\documentclass[fleqn,usenatbib]{mnras}

\usepackage{newtxtext,newtxmath}

\usepackage[T1]{fontenc}

\DeclareRobustCommand{\VAN}[3]{#2}
\let\VANthebibliography\thebibliography
\def\thebibliography{\DeclareRobustCommand{\VAN}[3]{##3}\VANthebibliography}

\usepackage{graphicx}	
\usepackage{amsmath}	

\usepackage{xcolor}

\title[Four New Triply Eclipsing Triples]{Four New Compact Triply Eclipsing Triples found with \textit{Gaia} and \textit{TESS}}

\author[D. R. Czavalinga et al.]{
Donát R. Czavalinga,$^{1,2}$\thanks{E-mail: czdonat@titan.physx.u-szeged.hu}
Tamás Borkovits,$^{1,2,3,4,5}$
Tibor Mitnyan,$^{1,2}$
Saul A. Rappaport,$^{6}$
András Pál $^{3}$
\\
$^{1}$ Baja Astronomical Observatory of University of Szeged, H-6500 Baja, Szegedi út, Kt. 766, Hungary\\
$^{2}$ HUN-REN-SZTE Stellar Astrophysics Research Group, H-6500 Baja, Szegedi út, Kt. 766, Hungary\\
$^{3}$ Konkoly Observatory, Research Centre for Astronomy and Earth Sciences, H-1121 Budapest, Konkoly Thege Miklós \'ut 15-17, Hungary\\
$^{4}$ ELTE Gothard Astrophysical Observatory, H-9700 Szombathely, Szent Imre h. u. 112, Hungary\\
$^{5}$ HUN-REN-ELTE Exoplanet Systems Research Group, H-9700 Szombathely, Szent Imre h. u. 112, Hungary \\
$^{6}$Department of Physics, Kavli Institute for Astrophysics and Space Research, M.I.T., Cambridge, MA 02139, USA
}

\date{Accepted XXX. Received YYY; in original form ZZZ}

\pubyear{2023}

\begin{document}
\label{firstpage}
\pagerange{\pageref{firstpage}--\pageref{lastpage}}
\maketitle

\begin{abstract}

This paper presents a comprehensive analysis of four triply eclipsing triple star systems, namely TIC\,88206187, TIC\,14839347, TIC\,298714297, and TIC\,66893949.  The four systems with third-body eclipses were found in the {\it TESS} lightcurves from among a sample of $\sim$400 matches between known eclipsing binaries and the \textit{Gaia} DR3 Non-Single Star \citep[NSS;][]{gaia22a,pourbaix22} solution database.  We combined photometric lightcurves, eclipse timing variations, archival spectral energy distributions, and theoretical evolution tracks in a robust photodynamical analysis to determine the orbital and system parameters. The triples have outer periods of 52.9, 85.5, 117, and 471 days, respectively.  All dozen stars have masses $\lesssim 2.6$\,M$_\odot$. The systems are quite flat with mutual inclination angles between the inner and outer orbital planes that are all $\lesssim 4^\circ$.  The outer mass ratios $(q \equiv M_3/M_{\rm bin})$ range from 0.39--0.76, consistent with our earlier collection of compact triply eclipsing triples.  TIC\,88206187  exhibits a fractional radius of the outer tertiary component $(r_B \equiv R_B/a_{\rm out})$ exceeding 0.1 (only the third such system known), and we consider its future evolution.  Finally, we compare our photodynamical analysis results and the orbital parameters given in the \textit{Gaia} DR3 NSS solutions, indicating decent agreement, but with the photodynamical results being more accurate.

\end{abstract}




\section{Introduction}

Triply eclipsing triple star systems are outstanding objects in which to study the properties and dynamics of multistellar systems. These are triple systems in which the inner binary eclipses the tertiary star, or vice versa, during the course of their outer orbit.  The outer orbital periods range from a month to a year or two, and therefore dynamical interactions can usually be observed within the duration of a PhD study.  Furthermore, even in the absence of radial velocity (RV) observations, readily available spaced-based photometric observations, such as with {\it Kepler} \citep{borucki10} or {\it TESS} \citep{ricker15} can yield precise eclipse timing measurements, as well as the photometry of the third-body eclipses.  These can be combined with the archival spectral energy distribution, and processed in a comprehensive spectro-photodynamical analysis to yield a full set of accurately determined stellar and orbital  parameters for the system.  In terms of the spatial configuration of the system, this analysis can provide such quantities as the ratio of orbital periods, the orbital eccentricities, and the mutual inclination of the two orbits.  In turn, these quantities play a crucial role in the long-term evolution of the whole system. These are therefore excellent laboratories for testing theories regarding the formation, evolution and final evolutionary states of multi-star systems.

The first such system was discovered in \textit{Kepler} data by \citet{carter11}, and the second one followed very shortly thereafter \citep{derekas11}. Within a few years, more than a dozen publications had reported the discovery of 15 more similar objects, among them the first detections using other observing facilities than \textit{Kepler}, including \textit{CoRoT} \citep{hajdu17} and OGLE \citep{hajdu22}. Since then, \textit{TESS} observations have accelerated the search for these objects and have led the way in such studies. \citet{borkovits20} reported the first identification and analysis of a triply eclipsing triple based on \textit{TESS} data, and in less than three years since then, 20 additional similar objects have been identified and analyzed in detail \citep{2020MNRAS.498.6034M,2022MNRAS.510.1352B,2022ApJ...938..133P,2022MNRAS.513.4341R,2023MNRAS.521..558R}. That means, thanks to \textit{Kepler} and then nowadays \textit{TESS}, the discovery of triply eclipsing triples has become more or less routine, and the numbers of known triply eclipsing triple systems have been steadily growing.

In our previous work \citep{czavalinga23}, we utilized the \textit{Gaia} Data Release 3 \citep[DR3;][]{gaia22b,2023A&A...674A..32B} Non-Single Star \citep[NSS;][]{gaia22a,pourbaix22} solutions to search for tertiary stars in eclipsing binary (EB) systems that had previously been reported in the literature. \textit{Gaia} DR3  provides a unique opportunity to search for close hierarchical triple stars across the entire sky based on long-term astrometric and spectroscopic observations. If we compare the eclipsing period with the period obtained from the \textit{Gaia} NSS solutions, and the period ratio is higher than 5, we can infer that the object is likely a triple-star system candidate with the inner period being that of the EB and the outer period of the tertiary star the one determined by \textit{Gaia}'s NSS solutions \citep[see][for details]{czavalinga23}. Using this method, we identified 403 potential compact hierarchical triple-star system candidates among $\sim$1 million known EBs, including four newly identified triply eclipsing triples. Here we provide a detailed analysis of the latter four systems. In Section \ref{sect:observations}, we describe the observational data and the methods utilized for the discovery and validation of the corresponding objects. In Section \ref{sec:photodynamical}, we construct a photodynamical model for each system, and then in Section \ref{sec:discussion}, we give a detailed discussion of the resulting models. In Section \ref{sect:comparison}, we compare the orbital parameters from our photodynamical models with those from the \textit{Gaia} NSS solutions, and finally we summarize our findings in Section \ref{sect:conclusions}.

\section{Discovery and Observations}
\label{sect:observations}

\subsection{\textit{TESS} observations} 
\label{sect:TESSobservation}
To validate the triple nature of our candidate systems we started by constructing a \textit{TESS} photometric lightcurve (LC) for each one. Our aim was to search for eclipse timing variations (ETVs) and possible third-body eclipses in the LC.  After downloading all the available Full-Frame Images (FFIs) from the MAST portal\footnote{\href{https://mast.stsci.edu/}{https://mast.stsci.edu/}}, we applied the same convolution-aided image subtraction photometry pipeline based on FITSH \citep{pal12} along with the same detrending process based on the {\sc W{\={o}}tan} \citep{hippke19} and lightkurve \citep{lightkurve18} Python packages that we used in our previous paper \citep[see details in][]{czavalinga23}.

\begin{figure*}
\begin{center}
\includegraphics[width=0.33 \textwidth]{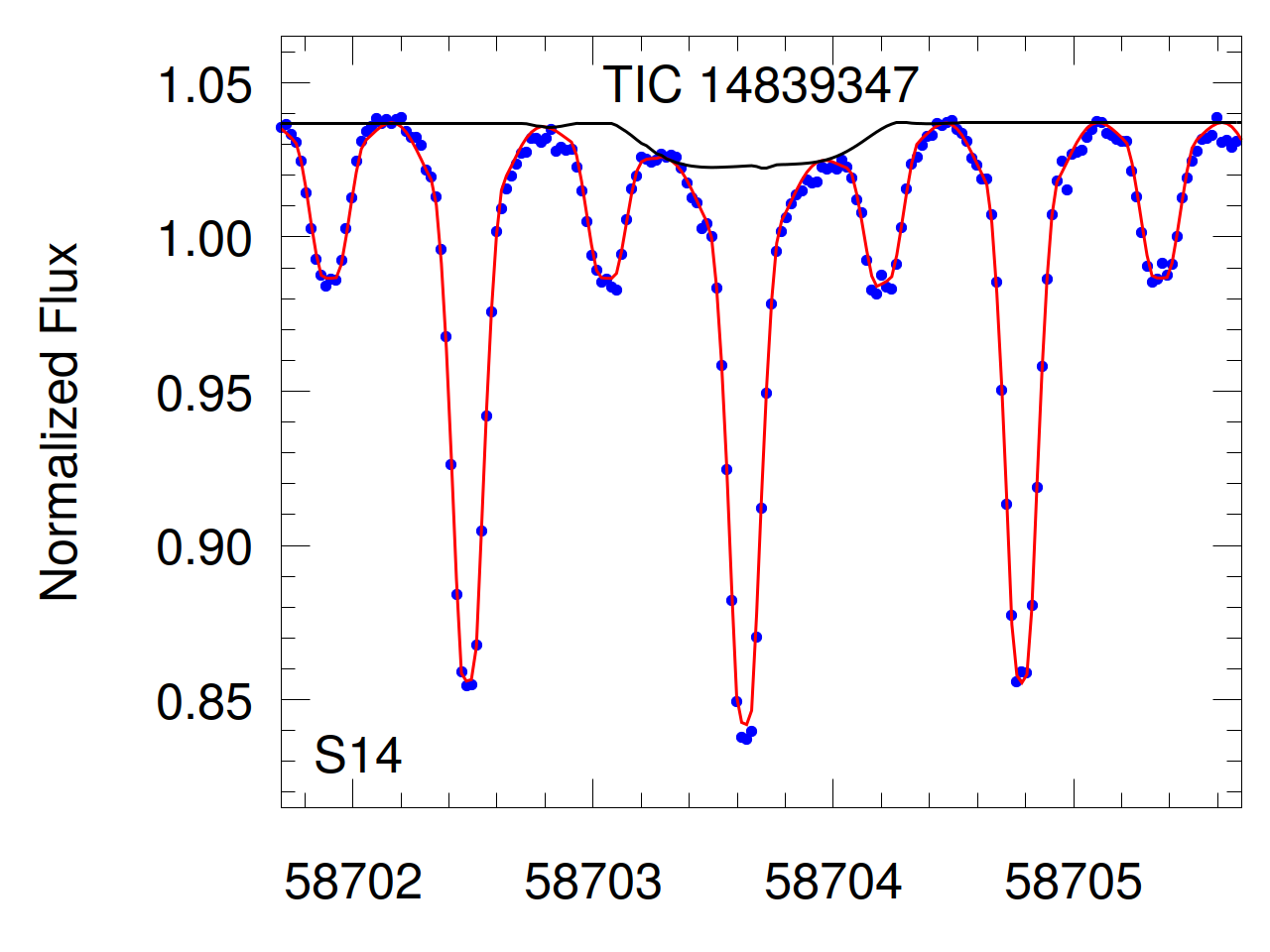} \includegraphics[width=0.33 \textwidth]{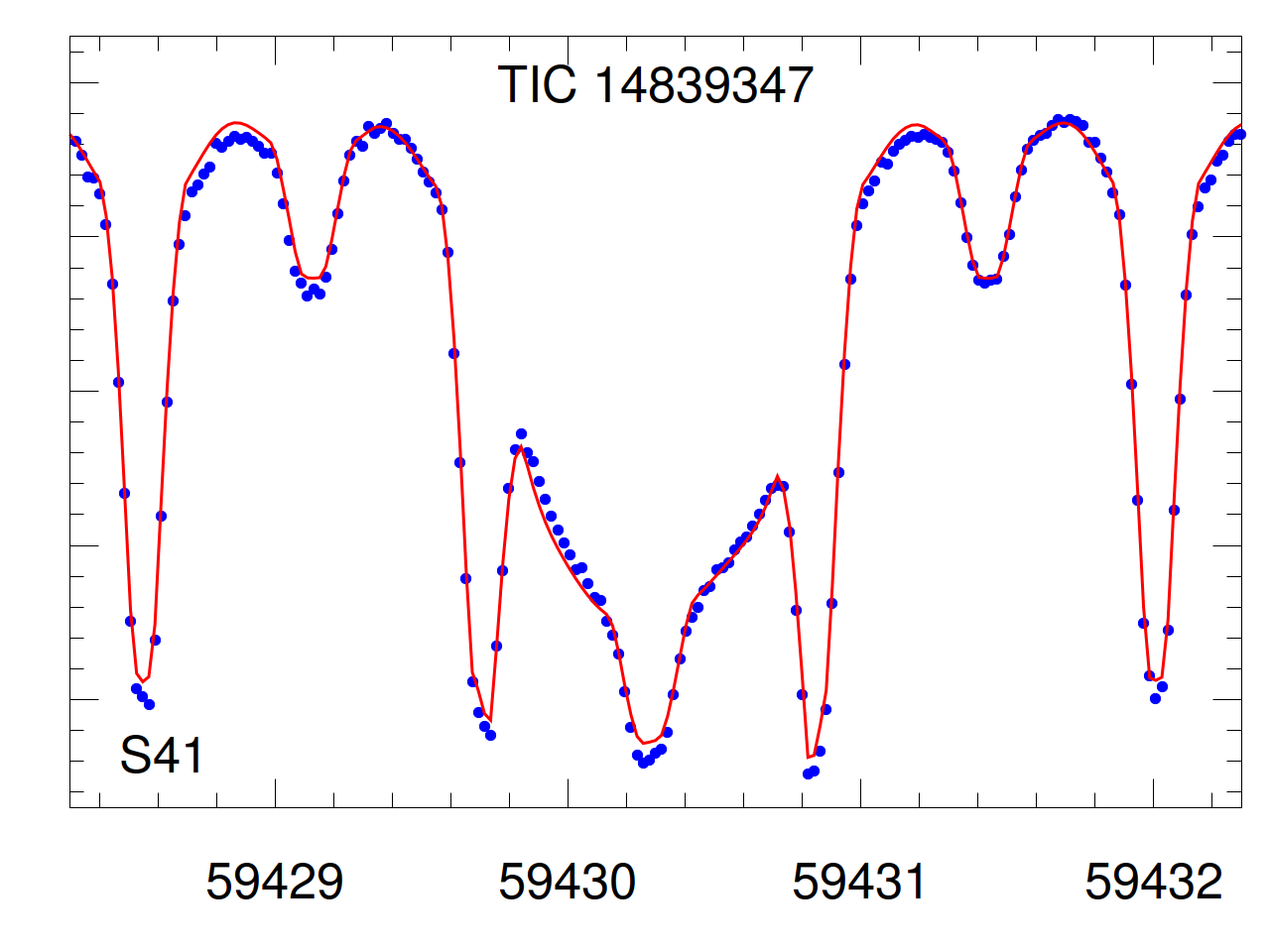} \includegraphics[width=0.33 \textwidth]{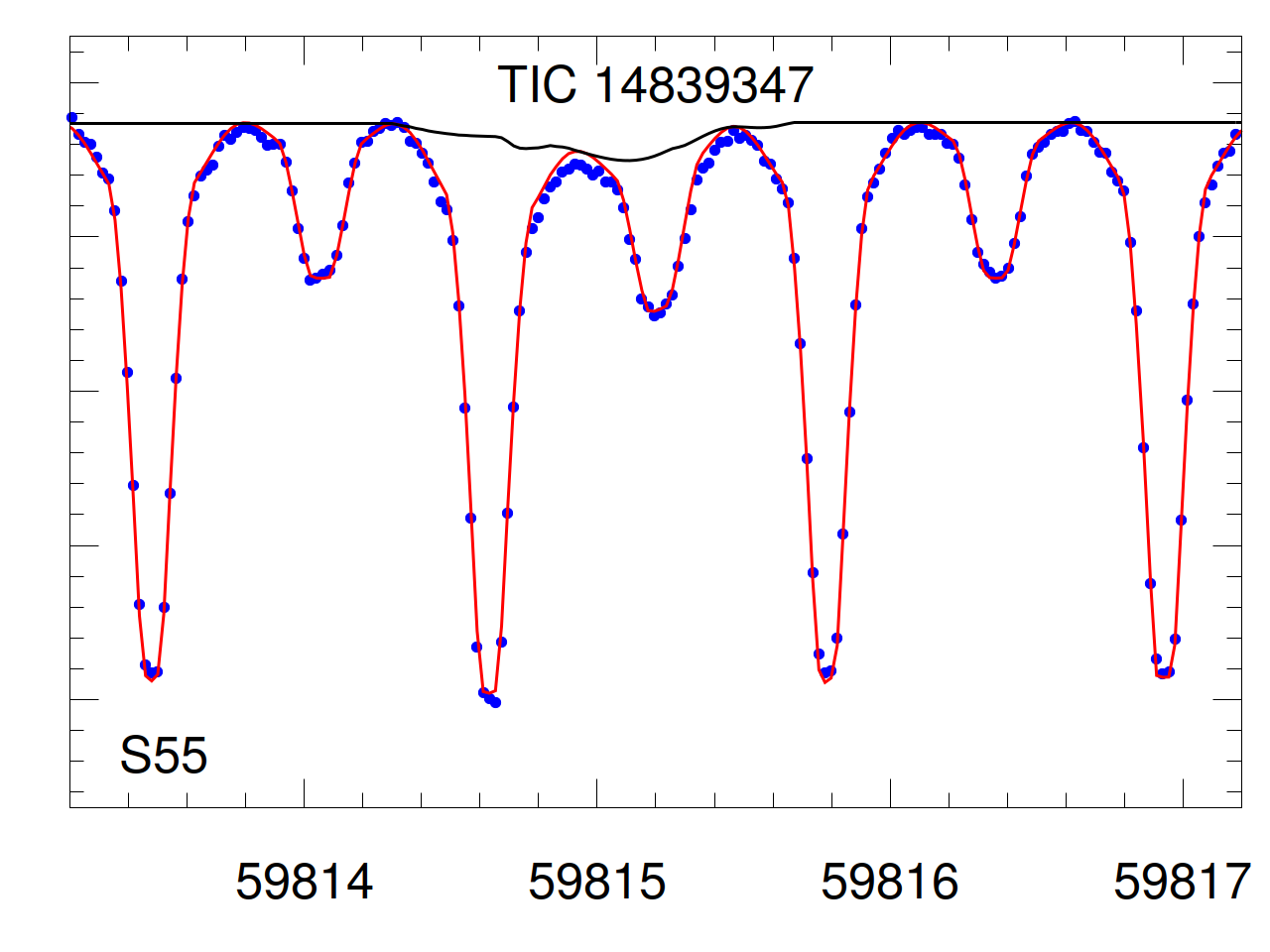} \vglue0.1cm
\includegraphics[width=0.43 \textwidth]{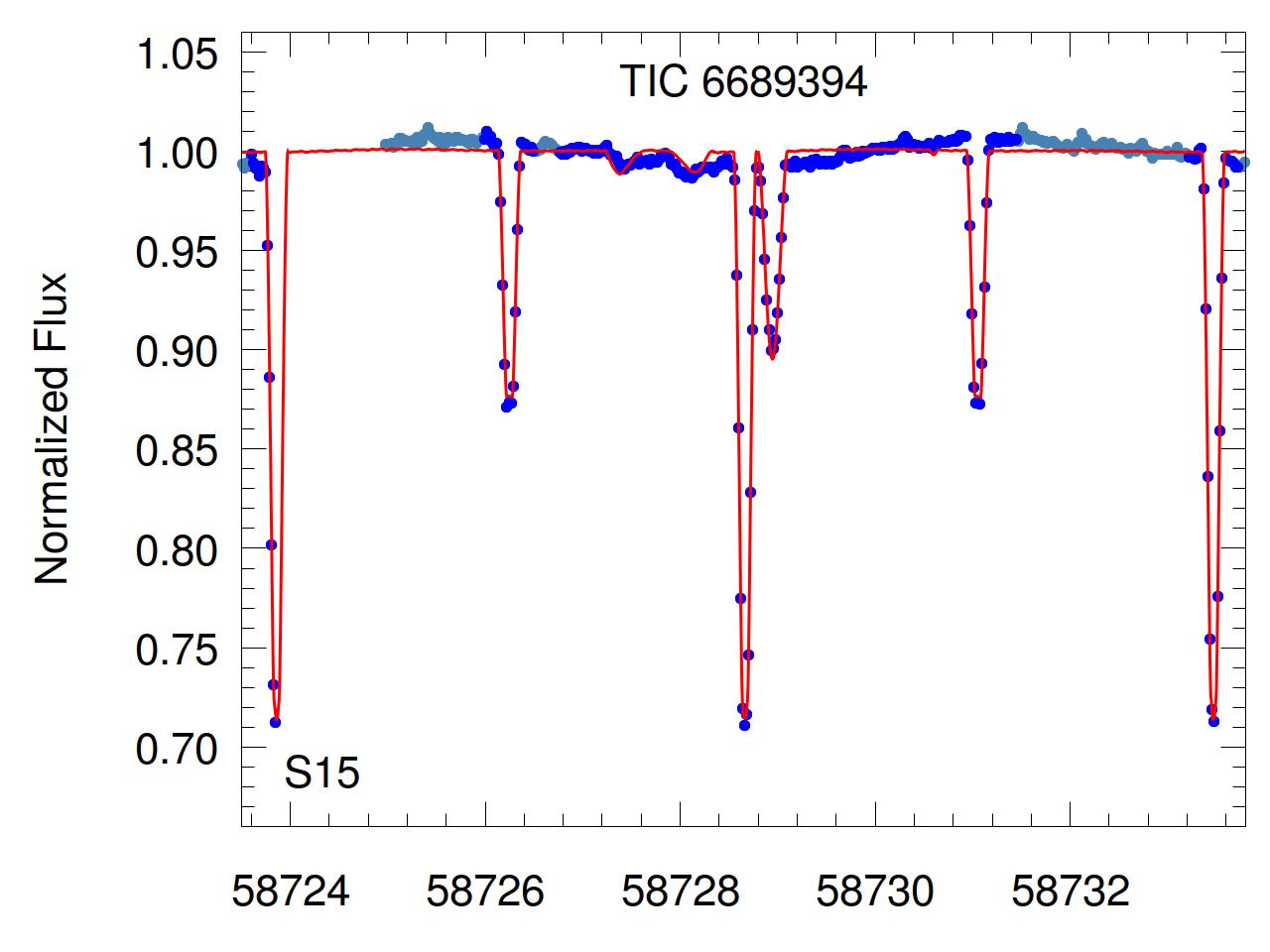}  \includegraphics[width=0.43 \textwidth]{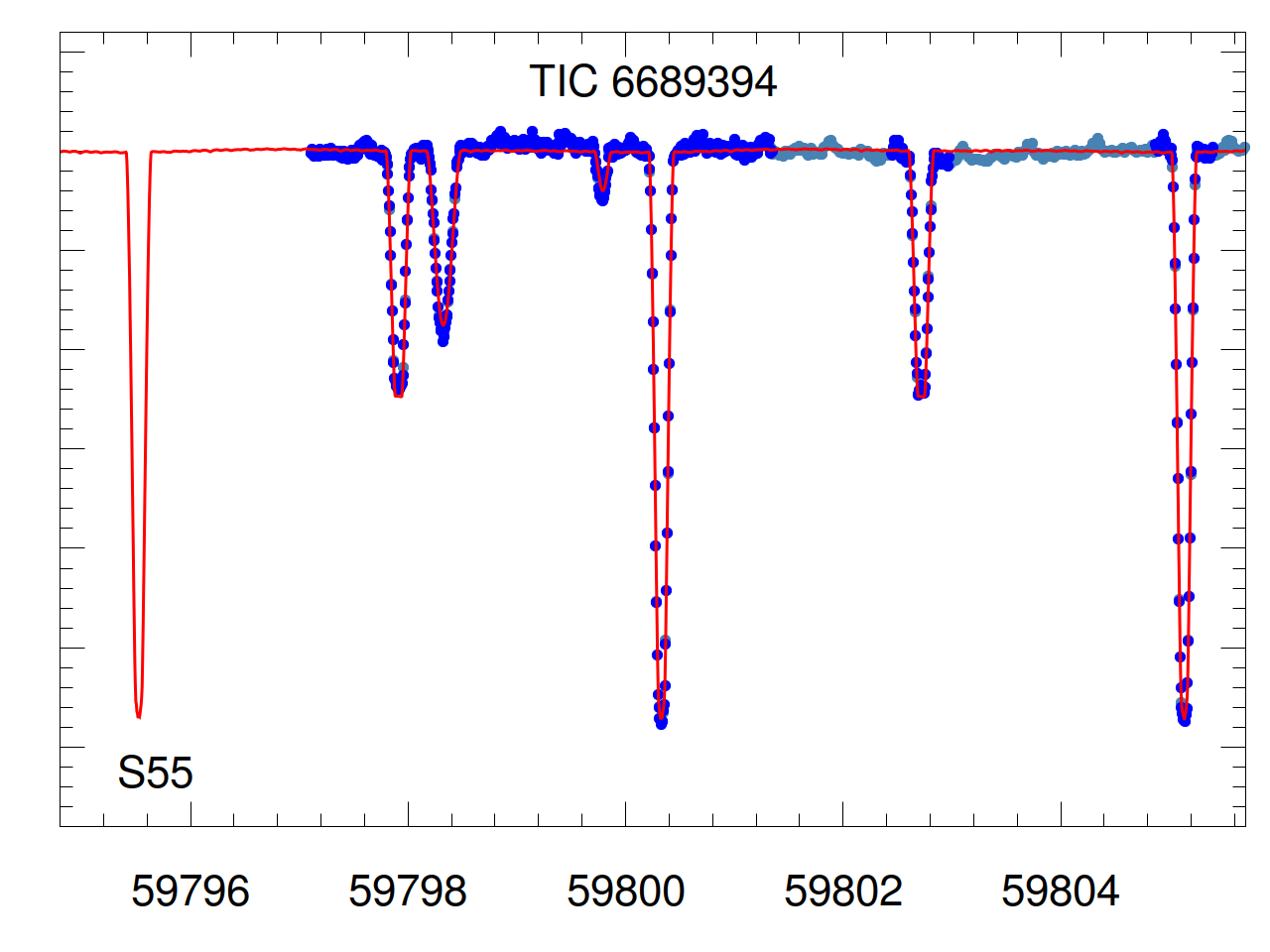} \vglue0.1cm  
\includegraphics[width=0.43 \textwidth]{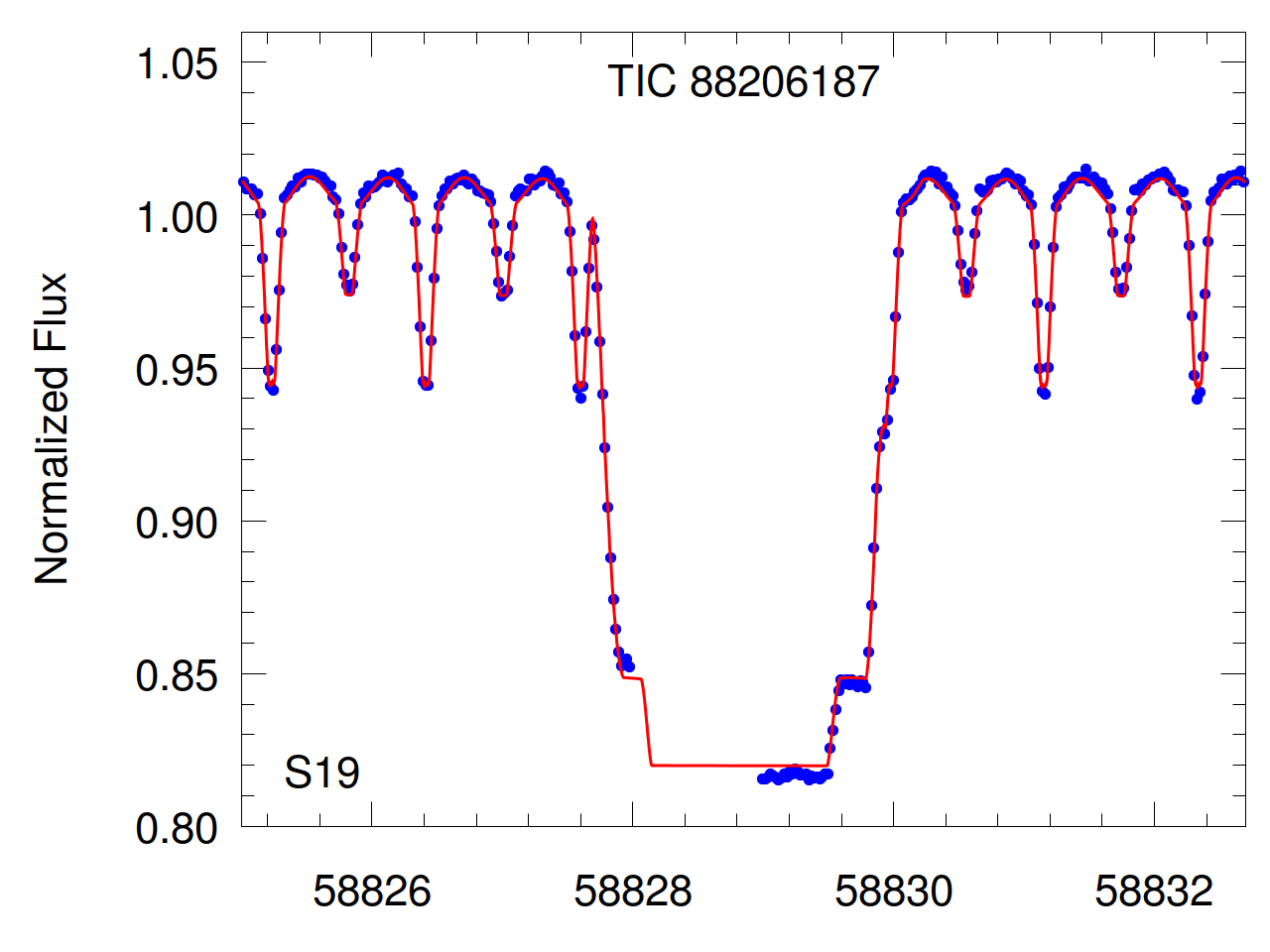}  \includegraphics[width=0.43 \textwidth]{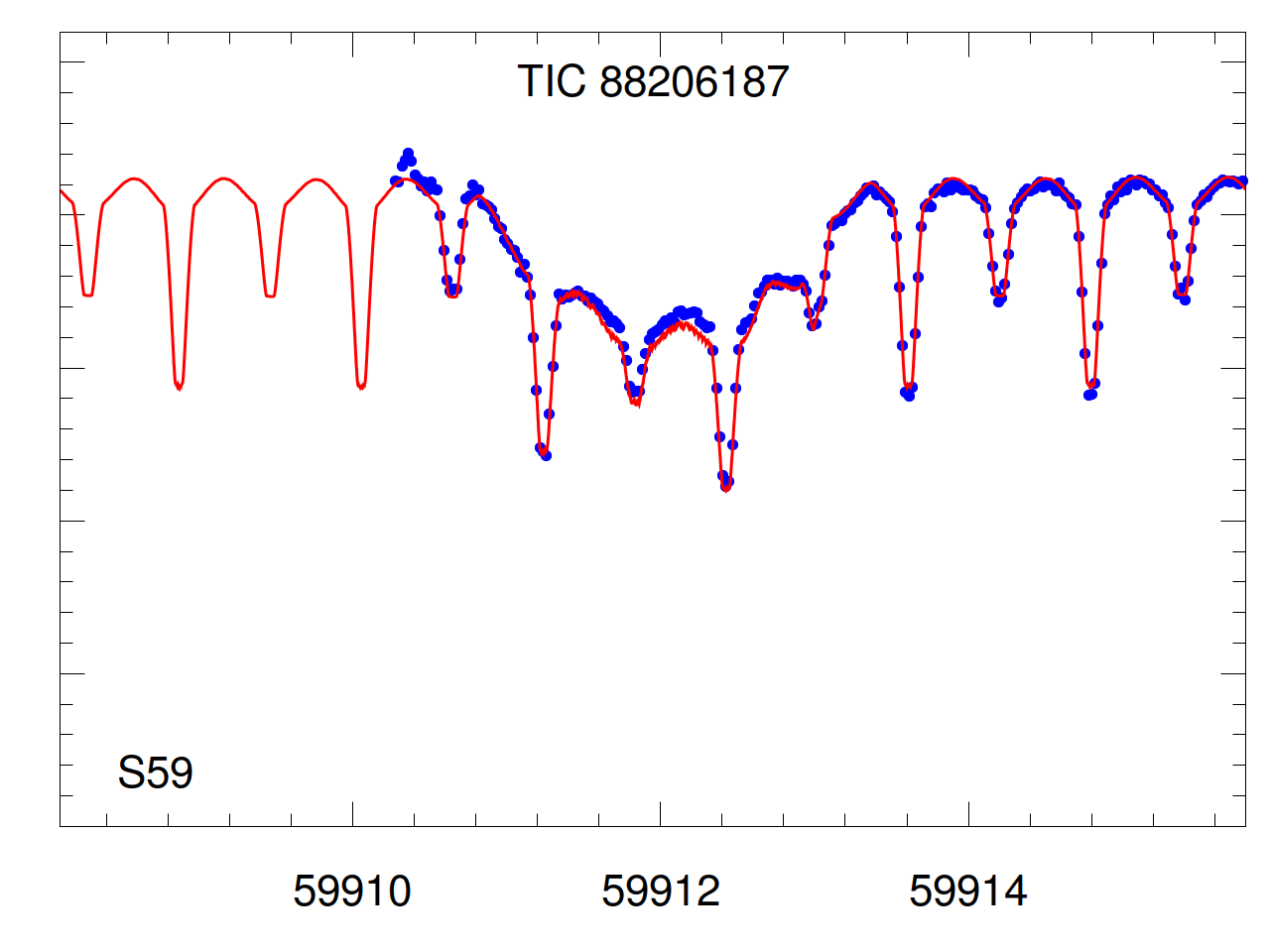} \vglue0.1cm  
\includegraphics[width=0.43 \textwidth]{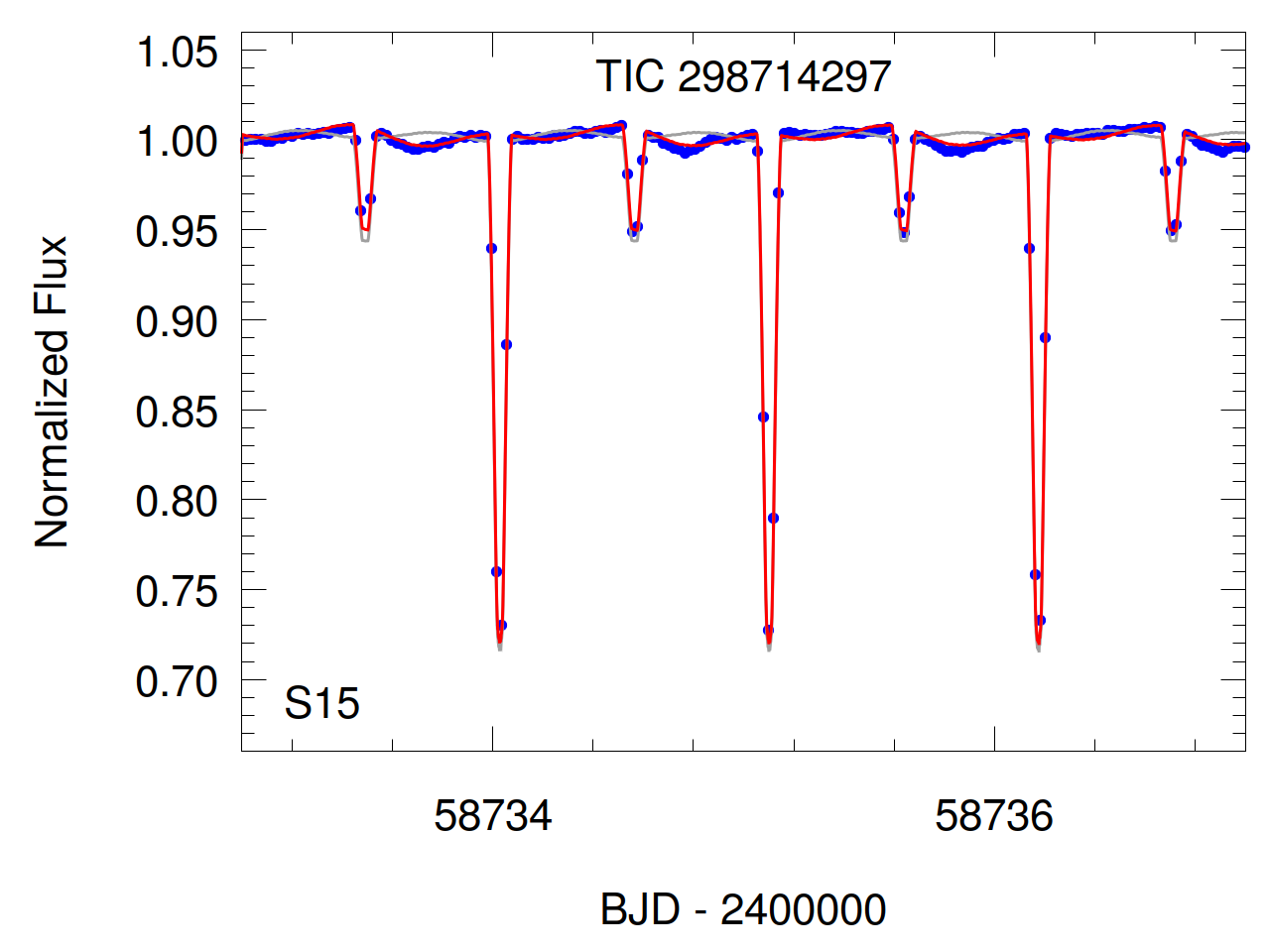}  \includegraphics[width=0.43 \textwidth]{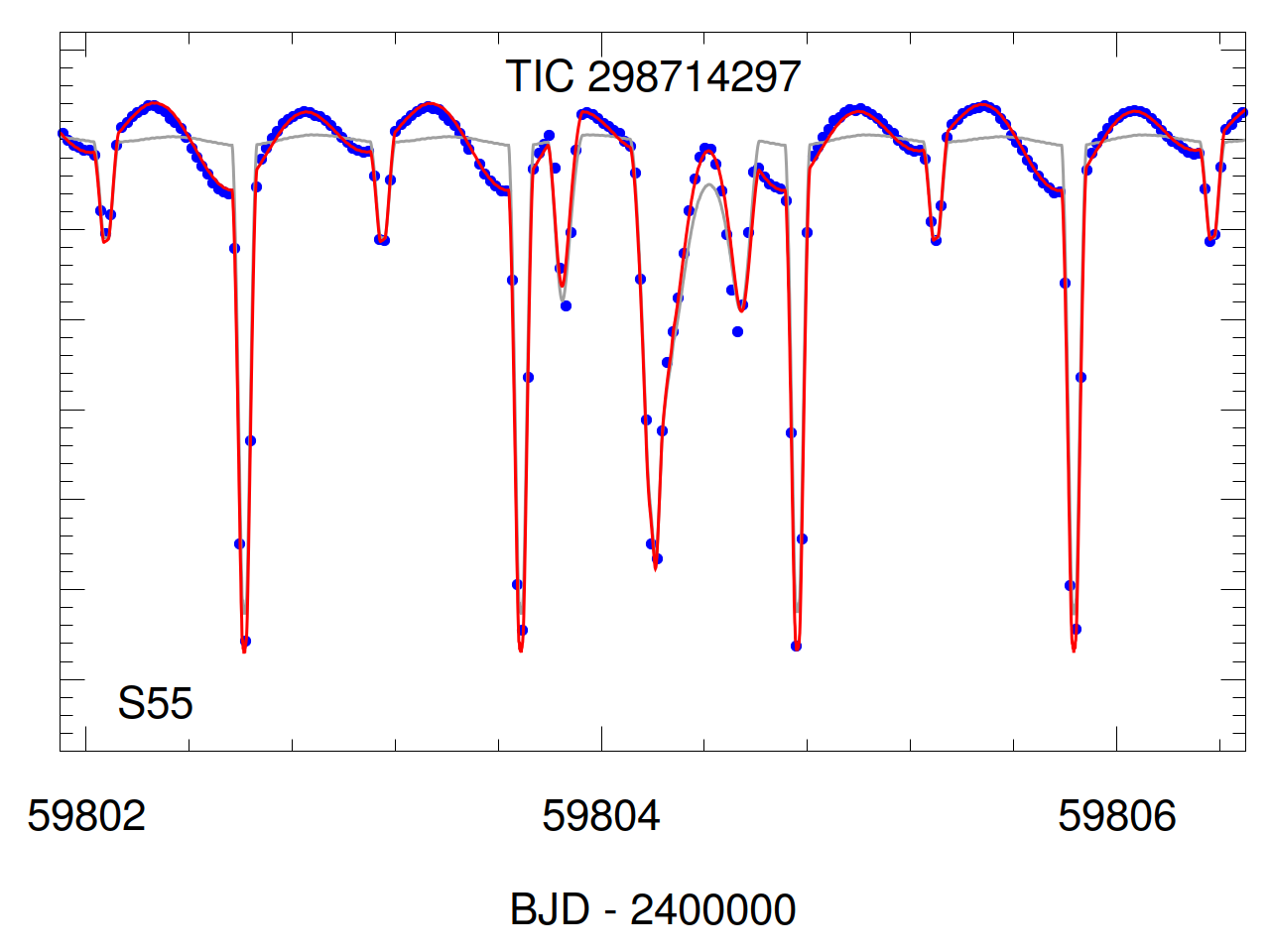}  

\caption{Sections of \textit{TESS} LCs with third-body eclipse events for each of the four triple systems The blue points are the \textit{TESS} observations, while the overplotted smooth red curves are the models from the photodynamical fit (see Sect.~\ref{sec:photodynamical}). In the left and right upper panels (the two very shallow secondary outer eclipses of TIC~14839347), the additional black lines represent just the flux occulations of the giant tertiary component as the EB passes in front of it. In the second row, the lighter blue points in the out-of-eclipse regions of TIC~66893949 were omitted from the photodynamical fits to save computation time. In the case of TIC 298714297 (bottom row) the gray curve displays the pure EB model (without the Fourier-modeled rotational distortions).}
\label{fig:triples}
\end{center}
\end{figure*}

We identified six LCs with potential extra third-body eclipses. Two of them had already been discovered by other teams, with one system being HD\,181068, the second triply eclipsing triple found in \textit{Kepler} observations \citep{derekas11}, and the other one being TIC\,229785001 recently analyzed in detail by \citet{2023MNRAS.521..558R}. The remaining four systems were completely unknown in the literature. These are as follows:

TIC\,14839347 was observed by \textit{TESS} in Sectors 14, 15, 41, and 55. We found a well-defined third-body eclipse in Sector 41, and then, after making use of the outer period from the \textit{Gaia} NSS solution, we additionally found two very shallow secondary eclipses in Sectors 14 and 55 (see upper left panel of Fig.~\ref{fig:triples}).

TIC\,66893949 was observed in Sectors 15, 41, and 55. In Sector 15, we identified an extra dip in the LC, which we interpreted as third-body eclipse. No eclipses occurred in Sector 41, but we observed two extra eclipsing events with different depths in Sector 55 (upper right panel of Fig.~\ref{fig:triples}). We note that recently \citet{rowan23} also independently discovered this object as a triply eclipsing triple, but they have not analyzed the system in any detail.

TIC\,88206187 was observed by \textit{TESS} in Sectors 19 and 59. We discovered a very deep and long, flat bottomed eclipse in the middle of Sector 19.  The source also produced another, similarly long in duration but much shallower, third-body eclipse in Sector 59 (lower left panel of Fig.~\ref{fig:triples}).

Finally, TIC\,298714297 was observed in Sectors 15, 55, and 56 by \textit{TESS}. We found evidence for one third-body eclipse in Sector 55, accompanied by what are likely prominent, stellar-activity induced variations in the LC (lower right panel of .Fig.~\ref{fig:triples}).

For the ETV calculation, we applied the same process as in our previous paper \citep{czavalinga23}, and that is described in detail in \citet{borkovits15}. In short, we fitted template polynomials to both the primary and secondary eclipses in the phase-folded, binned LC of each system. The mid-eclipse times were thereby determined for  each orbital cycle. We list the determined individual mid-eclipse times in Tables~\ref{Tab:TIC_014839347_ToM}--\ref{Tab:TIC_298714297_ToM} separately for all the four systems.

\subsection{ASAS-SN, ATLAS observations}
\label{Sect:ASASSN-ATLAS}
We looked up all four triply eclipsing triples in the ASAS-SN \citep{shappee14,kochanek17} and ATLAS \citep{tonry18,smith20} archives. All four systems have good data in the ASAS-SN archives, and two have useful data in the ATLAS archives (TIC\,298714297 and TIC\,66893949 are too bright for ATLAS). We readily see the EB LC in the archival data for all four systems.  

After we determine an accurate EB period for data that span of order a decade, we remove the EB LC by fitting for, and then subtracting out, between 50 and 100 orbital harmonics.  This allows for a more sensitive BLS search \citep[Box Least Squares;][]{kovacs02} for the eclipses of the outer orbit. 

\begin{figure}
\begin{center}
\includegraphics[width=0.99 \columnwidth]{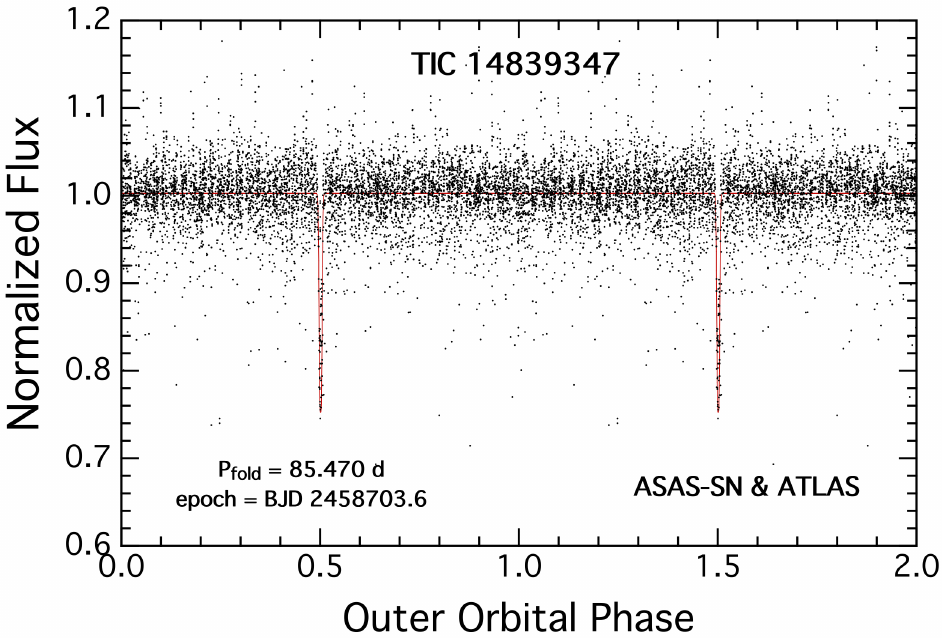}
\caption{Folded outer orbit of TIC\,14839347.}
\label{fig:TIC14839347archive}
\end{center}
\end{figure}

\begin{figure}
\begin{center}
\includegraphics[width=0.99 \columnwidth]{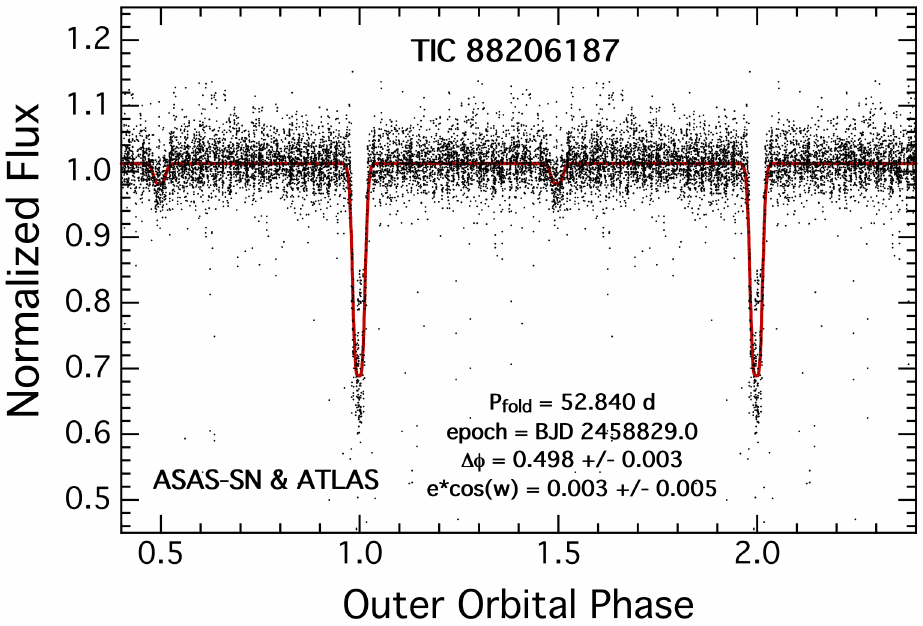}
\caption{Folded outer orbit of TIC\,88206187.}
\label{fig:TIC088206187archive}
\end{center}
\end{figure}

For TIC\,66893949 and TIC\,298714297 (with outer periods of 470.6 and 118.6 days, respectively) there are insufficient archival data, and too few outer eclipses observed, to make a statistically significant detection of the outer eclipses.  However, for TIC\,14839347 and TIC\,88206187, the outer periods of 85.47 days and  52.84 days, respectively, were robustly detected in a combination of the ASAS-SN and ATLAS data.  The results are shown in Figs.~\ref{fig:TIC14839347archive} and ~\ref{fig:TIC088206187archive}, respectively.  In the case of TIC\,88206187, Fig.~\ref{fig:TIC088206187archive} shows a shallow but clear secondary outer eclipse.  The phasing of the two outer eclipses yields $e_{\rm out} \cos \omega_{\rm out} = 0.003 \pm 0.005$, with an implication that the outer orbit is likely fairly circular.  

\begin{table*}
\caption{Main parameters of the four systems from the literature.}             
\label{tab:catalogdata}      
\centering                          
\begin{tabular}{c | c | c | c | c }        
Parameter & TIC\,14839347 & TIC\,66893949 & TIC\,88206187   & TIC\,298714297 \\
 \textit{Gaia} Source $\mathrm{ID^a}$ & 2058085351143518464 & 1870771652581668352 & 207943285475761280 &  1849392443551822848 \\
 \hline
$\mathrm{RA^a}$ [deg] & 306.408 & 311.711 & 82.739 & 324.463 \\
$\mathrm{DEC^a}$ [deg]& 37.895& 36.352 & 44.803   &  29.474  \\
$\mathrm{T^b [mag]}$ &12.2117 $\pm$ 0.0448 &11.6134 $\pm$ 0.0101&11.8518 $\pm$ 0.0078 &10.8136 $\pm$ 0.0069 \\
$\mathrm{G^a [mag]}$ &13.070626 $\pm$ 0.002975& 11.9993 ± 0.0002 & 12.478771  $\pm$  0.002864 &11.542984 $\pm$  0.003035 \\
$\mathrm{G_{bp}^a [mag]}$ &13.866420 $\pm$ 0.004369& 12.2891 $\pm$ 0.0008 & 13.006023  $\pm$  0.003710 & 12.203759 $\pm$  0.004678\\
$\mathrm{G_{rp}^a [mag]}$ &12.175914 $\pm$ 0.005246 & 11.5470 $\pm$ 0.0005 & 11.783351  $\pm$  0.004393 &10.743742 $\pm$  0.004989\\
$\mathrm{B^c [mag]}$ &14.837 $\pm$ 0.048& 12.47 $\pm$ 0.255& 13.403 $\pm$ 0.275 &13.309 $\pm$ 0.399\\
$\mathrm{V^c [mag]}$ &13.807 $\pm$ 0.149&12.082 $\pm$ 0.057&12.794 $\pm$ 0.069 &11.584 $\pm$ 0.138  \\
$\mathrm{g'^c [mag]}$ & 14.211$\pm$ 0.016 & 12.408$\pm$0.000 & 13.007 	$\pm$0.220 & 12.493$\pm$0.040 \\
$\mathrm{r'^c [mag]}$ &13.176$\pm$ 0.033 & 12.058$\pm$0.000 & 12.487$\pm$0.072 & 11.529$\pm$0.045 \\
$\mathrm{i'^c [mag]}$ &12.562$\pm$ 0.019 & 11.869$\pm$0.000 & 12.135$\pm$0.084 & 11.055$\pm$0.051 \\
$\mathrm{J^d [mag]}$ &10.81 $\pm$ 0.021&11.004 $\pm$ 0.021 & 10.828 $\pm$ 0.022 &9.717 $\pm$ 0.028\\
$\mathrm{H^d [mag]}$ &10.231 $\pm$ 0.018 &10.737 $\pm$ 0.017 & 10.351 $\pm$ 0.021 &9.092 $\pm$ 0.034\\
$\mathrm{K^d [mag]}$ &9.999 $\pm$ 0.016 &10.697 $\pm$ 0.016&10.206 $\pm$ 0.016 &8.965 $\pm$ 0.02 \\
$\mathrm{W1^e [mag]}$ &9.827 $\pm$ 0.023&10.685 $\pm$ 0.023 & 10.099 $\pm$ 0.023 &8.877 $\pm$ 0.023\\
$\mathrm{W2^e [mag]}$ &9.857 $\pm$ 0.02 &10.72 $\pm$ 0.021& 10.129 $\pm$ 0.02 &8.878 $\pm$ 0.02\\
$\mathrm{W3^e [mag]}$ &9.673 $\pm$ 0.088&10.666 $\pm$ 0.099&10.011 $\pm$ 0.062 &8.752 $\pm$ 0.024 \\
$\mathrm{W4^e [mag]}$ &7.836 $\pm$ 0.191&9.015 $\pm$ NaN & 8.67 $\pm$ 0.36 &8.82 $\pm$ 0.378\\
 
$\mathrm{T_{eff}^{a} [K]}$ & - & 5883 $\pm$ 27 & - &5498 $\pm$ 121\\
$\mathrm{Distance^f [pc]}$ & $1970_{-41}^{+48}$ &$587_{-7}^{+9}$ &$2490_{-78}^{+96}$ & $111.9_{-0.7}^{+0.7}$\\
$\mathrm{E[B-V]^c [mag]}$ &0.772 $\pm$ 0.065& 0.03929 $\pm$ 0.0124& 0.2585 $\pm$ NaN & 0.009 $\pm$ 0.00595  \\
$\mathrm{\mu_{\alpha}^a}$ &-2.657 $\pm$ 0.011   & 14.4171 $\pm$ 0.0202 &0.720 $\pm$  0.015 &-28.768 $\pm$  0.044\\
$\mathrm{\mu_{\delta}^a}$ &-7.034 $\pm$ 0.012& -15.2782 $\pm$ 0.0241 &-2.416 $\pm$  0.011 &-58.138 $\pm$  0.036\\
$\mathrm{RUWE^a}$ & 1.221	& 1.9273869 & 0.905 & 3.162 \\
NSS $\mathrm{model^a}$ & SB1 & Orbital & SB1 & Orbital \\
$\mathrm{P_{binary}^g [days]}$ & $1.154060_{-0.000052}^{+0.000050}$ & $4.805309_{-0.000005}^{+0.000004}$ & $1.184592_{-0.000063}^{+0.000055}$ & $1.072891_{-0.000019}^{+0.000010}$ \\
$\mathrm{P_{triple}^g [days]}$ &$85.530_{-0.017}^{+0.017}$&$471.03_{-0.07}^{+0.10}$&$52.922_{-0.039}^{+0.041}$&$117.24_{-0.31}^{+0.36}$ \\
\hline
\end{tabular}
\\
\textit{Notes.} '-' means that the value is not available (a) - \textit{Gaia} DR3 \citep{gaia22b}, coordinates: ICRS J2016.0; (b) - \textit{TESS} Input Catalog V8.2 \citep{paegert21}; (c) - AAVSO Photometric All Sky Survey (APASS) DR9 \citep{henden2015}; (d) - 2MASS All-Sky Catalog \citep{skrutskie2006}; (e) - ALLWISE Data Release \citep{allwise2014}; (f) - \citep{2021AJ....161..147B}; (g) - parameters from this paper
\end{table*}

\section{Photodynamical analysis}
\label{sec:photodynamical}

Similar to other \textit{TESS}-discovered triply eclipsing triple stars, we carried out a joint, simultaneous LC, ETV curve and spectral energy distribution (SED) analysis with the software package {\sc Lightcurvefactory}. The software package itself, and the consecutive steps of the complex analysis, were described in several papers, e.g., in \citet{2020MNRAS.493.5005B,2022MNRAS.510.1352B,2023MNRAS.521..558R} and, hence, we will not repeat the details here, but rather restrict ourselves to some particular notes about the current, specific systems.

Our analyses are mainly based on the \textit{TESS} LCs, which were processed in the manner described in Sect~\ref{sect:TESSobservation}. In contrast to the vast majority of our former analyses of triple and quadruple systems, in the case of 3 of the 4 systems reported here, we did not restrict our analyses only to the narrow regions of the inner and outer eclipses, dropping out the majority of the out-of-eclipse portions of the LCs, but rather we kept the complete \textit{TESS} time series.  The reason is that, apart from the somewhat wider 4.8-day inner EB in TIC~66893949, the other three, much more compact inner EBs, display significant ellipsoidal light variations in the out-of-eclipse LC sections, which we found to be worthwhile to retain for the analysis. Moreover, we note the special case of TIC~298714297, in which the LC exhibits further remarkable rotational variations with periods similar to the eclipsing period of the inner EB. We modelled these variations simultaneously with the full, eclipsing LC solution (but separately for Sectors 15 and 55-56), with the use of additional fits of harmonic functions in the manner described, e.g., in \citet{2018MNRAS.478.5135B,2021MNRAS.503.3759B}.

As usual, we fit the ETV curves of the four EBs simultaneously with their LCs. As our literature search did not turn up any additional previously recorded eclipse times, our ETV curves were restricted to only those times that we determined from the \textit{TESS} observations themselves. 

Due to the lack of publicly available RV timeseries for any of the four currently investigated systems, with the exception of TIC\,14839347, we used a combination of (i) observed composite SED values (tabulated from the available catalog passband magnitudes, and listed in Table~\ref{tab:catalogdata}) and (ii) pre-computed, tabulated \texttt{PARSEC} isochrones \citep{2012MNRAS.427..127B} to find stellar masses and corresponding effective temperatures, as well. The use of \texttt{PARSEC} isochrone-based SED analysis for this purpose was explained in \citealt{2020MNRAS.493.5005B}, while a comparison of the accuracy and efficiency of such an astrophysical model-dependent analysis with the classical, astrophysical model-independent analysis (the latter of which is based on RV data), was carried out and discussed in \citealt{2022MNRAS.510.1352B}.

In the case of one of the four systems, TIC\,14839347, our preliminary analysis run, however, revealed a situation where the secondary is fairly low in mass, but large enough to fill or nearly fill its Roche lobe. 
We therefore suspect that the secondary star in TIC\,14839347 may have already transferred a portion of its envelope to the (current) primary star. As a consequence, for this triple we cannot use this proxy method because the precomputed \texttt{PARSEC} evolution grids are valid only for single stars, i.e., for such binary components which have not previously exchanged mass. 
Therefore, in the case of this particular triple, we followed an iterative method as follows. First, we carried out {\sc Lightcurvefactory} fitting, excluding the SED analysis section. In such a way, we obtained strong constraints for the relative (or, fractional) radii of the three stars, as well as their temperature ratios. Then, using these relative quantities, as well as the \textit{Gaia} EDR3 derived distance \citep{2021AJ....161..147B}, we made an independent SED analysis without the use of any astrophysical preassumptions, with a slightly modified version of the method described in \citet{2022MNRAS.513.4341R}, Sect. 3. In this way, we obtained likely mass and temperature ranges for the three constituent stars. Then we used the SED-determined values for the mass of the primary of the inner binary ($m_\mathrm{Aa}$) and the effective temperature of the tertiary ($T_\mathrm{B}$) with their statistical uncertainties as Gaussian priors, and reiterated the whole simulatneous LC and ETV curve fitting procedure with {\sc Lightcurvefactory}. With this iterative procedure, we were able to find the system geometry and dynamics robustly, while also inferring the physical properties of the constituent stars at a reasonable level of accuracy.

In regard to TIC\,14839347, we also had to introduce some further modification to the light-curve fitting part of the complex analysis. We found that, in this case, the reflection/irradiation effect gives a significant contribution to the LC of the inner EB. Hence, we ``switched on'' this effect for the modeling. First we set the bolometric albedos according to the theoretically expected values of $A=1.0$ and $0.5$ for the primary (Aa) and secondary stars (Ab), (i.e., radiative and convective envelopes) respectively. We found, however, that the fit becomes distinctly better when one sets $A_\mathrm{Ab}\approx0.8$. For this reason, finally we decided to adjust, as an exception to our usual approach, the bolometric albedos and gravity darkening exponents ($\beta$) for these two stars. 

Finally we note the fact that none of the four investigated triples is a `tight' system (the outer-to-inner period ratios, $P_\mathrm{out}/P_\mathrm{in}$ are somewhat large, ranging from 45 for TIC~88206187, to 109 TIC~298714297) and, hence, one cannot expect significant third-body perturbations which would produce significant departures from simple Keplerian motions. On the other hand, considering the fact that the third-body eclipses are extremely sensitive to the current system geometry, similar to our former analyses (where much tighter triples were analysed) we integrated the motion of the three stellar components numerically, instead of approximating the motion with the superposition of two Keplerian orbits. The numerical integrator built into {\sc Lightcurvefactory} is a seventh order Runge-Kutta-Nystrom integrator. This integrator provides the gravitationally (pure Newtonian two- and three-body terms), tidally (within the framework of the equilibrium tide model) and, optionally, the relativistically perturbed Cartesian Jacobian coordinates (and velocities) for each observational instant. A detailed description of the integrator itself, and its implementation in the photodynamical software package can be found in \citet{2004A&A...426..951B,2019MNRAS.483.1934B,2019MNRAS.487.4631B}.

We initialized and ran several MCMC chains for each of the four triple systems.  The best fit parameters (median values of the posteriors), their $1-\sigma$ uncertainties, as well as several derived parameters, are tabulated in Tables~\ref{tab:syntheticfit_TIC014839347066893949} and \ref{tab:syntheticfit_TIC088206187298714297}. Here we note, that while most of the given derived parameters (e.g., the semi-major axes, radial velocity amplitudes, bolometric lumonisities, etc.) do not require any further explanations, we discuss briefly the calculations and significances of the different apsidal and nodal motion parameters, tabulated in between the orbital elements and the stellar parameters.

We give the theoretical apsidal motion periods both in the observational and the dynamical frames of reference ($P_\mathrm{apse}$ and $P_\mathrm{apse}^\mathrm{dyn}$). While the formulae for their calculations, from the initial osculating orbital elements at each accepted MCMC trial step, are discussed in detail in Sect.~6.3 of \citet{2021ApJ...917...93K}, here we emphasize only the fact that the observational and dynamical apsidal motions and, hence, their timescales are substantially different. From an observational point of view, what is significant is the revolution of the argument of periastron in the observational frame of reference.\footnote{The basic plane of the observational reference frame is the tangential plane of the sky, which is perpendicular to the line of sight to the target as seen by the observer. The argument of pericenter in this frame is the angle measured from one of the intersections of the orbit and the nodal line of this tangential plane and the orbital plane, to the pericenter point, along the orbit.} This revolution manifests itself, e.g., in the periodic, quasi-sinusoidal and anticorrelated shifts of the primary and secondary eclipses, in the case of eccentric EBs. In contrast to this, the revolution of the argument of pericenter in the dynamical frame\footnote{The basic plane of the dynamical reference frame is the invariable plane of the triple system, i.e., the plane whose normal is parallel to the constant total angular momentum of the triple. The argument of pericenter in this frame is measured from one of the intersections of the orbit and the nodal line of this invariable plane and the orbital plane, to the pericenter point, along the orbit.} cannot be observed directly, however, this dynamical argument of periastron and, hence, the dynamical-frame apsidal motion are significant for dynamical studies, as this parameter is one that occurs in the perturbation equations. The observable-frame apsidal motion, in general, is a non-linear combination of the dynamical-frame apsidal motion and the (dynamical) nodal regression, and hence, their periods (i.e., $P_\mathrm{apse}$ and $P_\mathrm{apse}^\mathrm{dyn}$) may be substantially different (see, e.g., \citealt{borkovits15} for further details). Note also, that besides the apsidal motion and nodal regression ($P_\mathrm{node}^\mathrm{dyn}$) periods, we  give separately the three components of the apsidal advance rates (classical tidal, $\Delta\omega_{\rm tide}$; general relativistic, $\Delta\omega_{\rm GR}$; and dynamical third-body, $\Delta\omega_{\rm 3b}$, respectively) for one orbital revolution of the inner and outer orbits, respectively. These contributions are calculated in the dynamical frame of reference.

Finally, note that the third-body eclipse sections of the model LCs from the best-fit complex photodynamical solutions are plotted in Fig.~\ref{fig:triples}, while the model ETV curves of the best-fit solutions are shown in Figs.~\ref{fig:T014839347ETV}--\ref{fig:T298714297ETV}. We briefly discuss our results for each system separately in Sect.~\ref{sec:discussion}.

\begin{table*}
 \centering
\caption{Orbital and astrophysical parameters of TICs 14839347 and 66893949 from the joint photodynamical \textit{TESS}, ETV, SED and \texttt{PARSEC} isochrone solution. Note that the orbital parameters are instantaneous, osculating orbital elements and are given for epoch $t_0$ (first row).  }
 \label{tab:syntheticfit_TIC014839347066893949}
\begin{tabular}{@{}lllllll}
\hline
& \multicolumn{3}{c}{TIC\,14839347} & \multicolumn{3}{c}{TIC\,66893949} \\
\hline
\multicolumn{7}{c}{orbital elements} \\
\hline
   & \multicolumn{3}{c}{subsystem} & \multicolumn{3}{c}{subsystem} \\
   & \multicolumn{2}{c}{Aa--Ab} & A--B & \multicolumn{2}{c}{Aa--Ab} & A--B \\
  \hline
  $t_0$ [BJD - 2400000] & \multicolumn{3}{c}{58\,683.0} & \multicolumn{3}{c}{58\,711.0}  \\
  $P$ [days] & \multicolumn{2}{c}{$1.154060_{-0.000052}^{+0.000050}$} & $85.530_{-0.017}^{+0.017}$ & \multicolumn{2}{c}{$4.805309_{-0.000005}^{+0.000004}$} & $471.03_{-0.07}^{+0.10}$  \\
  $a$ [R$_\odot$] & \multicolumn{2}{c}{$6.769_{-0.034}^{+0.043}$} & $144.2_{-0.9}^{+1.7}$ & \multicolumn{2}{c}{$16.29_{-0.13}^{+0.16}$} & $386.4_{-3.4}^{+4.2}$ \\
  $e$ & \multicolumn{2}{c}{$0.0005_{-0.0003}^{+0.0005}$} & $0.042_{-0.013}^{+0.013}$ & \multicolumn{2}{c}{$0.0050_{-0.0003}^{+0.0013}$} & $0.4016_{-0.0038}^{+0.0039}$ \\
  $\omega$ [deg] & \multicolumn{2}{c}{$147_{-97}^{+138}$} & $269.6_{-3.1}^{+2.8}$ & \multicolumn{2}{c}{$317_{-35}^{+32}$} & $25.8_{-1.5}^{+1.3}$ \\ 
  $i$ [deg] & \multicolumn{2}{c}{$88.75_{-1.13}^{+1.13}$} & $86.51_{-0.13}^{+0.12}$ & \multicolumn{2}{c}{$90.18_{-0.29}^{+0.34}$} & $90.222_{-0.009}^{+0.010}$  \\
  $\mathcal{T}_0^\mathrm{inf}$ [BJD - 2400000]$^a$ & \multicolumn{2}{c}{$58\,684.0427_{-0.0001}^{+0.0001}$} & ${59\,430.6949_{-0.0210}^{+0.0213}}$ & \multicolumn{2}{c}{$58\,714.2412_{-0.0002}^{+0.0002}$} & ${58\,729.4535_{-0.0245}^{+0.0225}}$  \\
  $\tau$ [BJD - 2400000]$^b$ & \multicolumn{2}{c}{$58\,683.56_{-0.46}^{+0.20}$} & $59\,429.6_{-0.7}^{+0.8}$ & \multicolumn{2}{c}{$58\,710.46_{-0.42}^{+0.38}$} & $58\,349.1_{-1.6}^{+1.4}$  \\
  $\Omega$ [deg] & \multicolumn{2}{c}{$0.0$} & $-2.9_{-3.1}^{+1.7}$ & \multicolumn{2}{c}{$0.0$} & $0.60_{-0.48}^{+0.53}$ \\
  $i_\mathrm{mut}$ [deg] & \multicolumn{3}{c}{$3.5_{-1.0}^{+3.3}$} & \multicolumn{3}{c}{$0.69_{-0.35}^{+0.49}$} \\
  \hline
  mass ratio $[q=m_\mathrm{sec}/m_\mathrm{pri}]$ & \multicolumn{2}{c}{$0.250_{-0.015}^{+0.016}$} & $0.757_{-0.019}^{+0.050}$ & \multicolumn{2}{c}{$0.670_{-0.005}^{+0.005}$} & $0.388_{-0.006}^{+0.007}$ \\
$K_\mathrm{pri}$ [km\,s$^{-1}$] & \multicolumn{2}{c}{$59.4_{-3.1}^{+3.3}$} & $36.7_{-0.7}^{+1.7}$ & \multicolumn{2}{c}{$68.84_{-0.40}^{+0.44}$} & $12.67_{-0.23}^{+0.28}$ \\ 
$K_\mathrm{sec}$ [km\,s$^{-1}$] & \multicolumn{2}{c}{$237.5_{-2.3}^{+2.1}$} & $48.4_{-0.8}^{+0.5}$ & \multicolumn{2}{c}{$102.80_{-1.06}^{+1.30}$} & $32.67_{-0.24}^{+0.27}$ \\ 
  \hline  
\multicolumn{7}{c}{Apsidal and nodal motion related parameters$^c$} \\
\hline  
$P_\mathrm{apse}$ [year] & \multicolumn{2}{c}{$0.80_{-0.02}^{+0.02}$} & $889_{-33}^{+33}$ & \multicolumn{2}{c}{$417_{-6}^{+6}$} & $2821_{-29}^{+30}$ \\
$P_\mathrm{apse}^\mathrm{dyn}$ [year] & \multicolumn{2}{c}{$0.79_{-0.02}^{+0.02}$} & $48.0_{-1.6}^{+0.6}$ & \multicolumn{2}{c}{$204_{-3}^{+2}$} & $350_{-3}^{+3}$\\
$P_\mathrm{node}^\mathrm{dyn}$ [year] & \multicolumn{3}{c}{$50.7_{-0.7}^{+1.8}$} & \multicolumn{3}{c}{$399_{-4}^{+5}$} \\
$\Delta\omega_\mathrm{3b}$ [arcsec/cycle] & \multicolumn{2}{c}{$156_{-3}^{+7}$} & $6321_{-79}^{+224}$ & \multicolumn{2}{c}{$79.4_{-0.9}^{+1.0}$} & $4779_{-45}^{+47}$ \\
$\Delta\omega_\mathrm{GR}$ [arcsec/cycle] & \multicolumn{2}{c}{$3.81_{-0.04}^{+0.05}$} & $0.315_{-0.004}^{+0.008}$  & \multicolumn{2}{c}{$1.26_{-0.05}^{+0.04}$} & $0.088_{-0.004}^{+0.003}$ \\
$\Delta\omega_\mathrm{tide}$ [arcsec/cycle] & \multicolumn{2}{c}{$5014_{-131}^{+123}$} &$1.7_{-0.1}^{+0.1}$ & \multicolumn{2}{c}{$2.93_{-0.12}^{+0.15}$} & $0.00102_{-0.00004}^{+0.0005}$ \\
\hline
\multicolumn{7}{c}{stellar parameters} \\
\hline
   & Aa & Ab &  B & Aa & Ab &  B \\
  \hline
 \multicolumn{7}{c}{Relative quantities and atmospheric properties} \\
  \hline
 fractional radius [$R/a$] & $0.4239_{-0.0041}^{+0.0037}$ & $0.2676_{-0.0045}^{+0.0046}$ & $0.0573_{-0.0016}^{+0.0018}$ & $0.0949_{-0.0006}^{+0.0007}$ & $0.0549_{-0.0007}^{+0.0009}$ & $0.00223_{-0.00006}^{+0.00008}$ \\
 temperature relative to $(T_\mathrm{eff})_\mathrm{Aa}$ & $1$ & $0.548_{-0.022}^{+0.016}$ & $0.559_{-0.020}^{+0.024}$ & $1$ & $0.803_{-0.006}^{+0.006}$ & $0.786_{-0.009}^{+0.008}$ \\
 fractional flux [in \textit{TESS}-band] & $0.3703_{-0.0125}^{+0.0122}$ & $0.0372_{-0.0015}^{+0.0012}$ & $0.5352_{-0.0265}^{+0.0478}$ & $0.7644_{-0.0140}^{+0.0107}$ & $0.1230_{-0.0016}^{+0.0016}$ & $0.1039_{-0.0082}^{+0.0011}$ \\
 grav. dark. exponent [$\beta$]$^d$ & $0.41_{-0.25}^{+0.39}$ & $0.17_{-0.13}^{+0.20}$ &  ... & ... & ... & ... \\ 
 albedo [$A$]$^d$ & $0.97_{-0.06}^{+0.05}$ & $0.76_{-0.05}^{+0.05}$ & ... & ... & ... & ... \\
 \hline
 \multicolumn{7}{c}{Physical Quantities} \\
  \hline 
 $m$ [M$_\odot$] & ${2.497_{-0.025}^{+0.030}}^d$ & $0.625_{-0.038}^{+0.041}$ & $2.359_{-0.061}^{+0.166}$ & $1.503_{-0.038}^{+0.050}$ & $1.007_{-0.021}^{+0.026}$ & $0.974_{-0.034}^{+0.043}$ \\
 $R$ [R$_\odot$] & $2.871_{-0.032}^{+0.031}$ & $1.812_{-0.038}^{+0.039}$ & $8.242_{-0.191}^{+0.313}$ & $1.547_{-0.023}^{+0.023}$ & $0.894_{-0.018}^{+0.023}$ & $0.861_{-0.029}^{+0.039}$  \\
 $T_\mathrm{eff}$ [K] & $8977_{-371}^{+380}$ & $4906_{-108}^{+95}$ & ${5019_{-44}^{+40}}^c$ & $6756_{-157}^{+147}$ & $5421_{-90}^{+81}$ & $5298_{-119}^{+150}$  \\
 $L_\mathrm{bol}$ [L$_\odot$] & $48.0_{-6.9}^{+8.2}$ & $1.71_{-0.14}^{+0.13}$ & $38.7_{-2.6}^{+3.7}$ & $4.47_{-0.49}^{+0.55}$ & $0.618_{-0.059}^{+0.074}$ & $0.523_{-0.077}^{+0.117}$ \\
 $M_\mathrm{bol}$ & $0.54_{-0.17}^{+0.17}$ & $4.16_{-0.08}^{+0.09}$ & $0.77_{-0.10}^{+0.14}$ & $3.14_{-0.13}^{+0.13}$ & $5.29_{-0.12}^{+0.11}$ & $5.47_{-0.22}^{+0.17}$ \\
 $M_V           $ & $0.60_{-0.11}^{+0.12}$ & $4.50_{-0.12}^{+0.15}$ & $1.06_{-0.10}^{+0.08}$ & $3.09_{-0.13}^{+0.13}$ & $5.42_{-0.15}^{+0.14}$ & $5.65_{-0.27}^{+0.22}$ \\
 $\log g$ [dex] & $3.920_{-0.008}^{+0.009}$ & $3.718_{-0.009}^{+0.009}$ & $2.978_{-0.025}^{+0.034}$ & $4.235_{-0.004}^{+0.005}$ & $4.537_{-0.011}^{+0.008}$ & $4.555_{-0.020}^{+0.015}$ \\
 \hline
\multicolumn{7}{c}{Global system parameters} \\
  \hline
$\log$(age) [dex] & \multicolumn{3}{c}{$-$} & \multicolumn{3}{c}{$8.77_{-0.21}^{+0.12}$} \\
$[M/H]$  [dex]    & \multicolumn{3}{c}{$-$} & \multicolumn{3}{c}{$0.36_{-0.06}^{+0.04}$} \\
$E(B-V)$ [mag]    & \multicolumn{3}{c}{$-$} & \multicolumn{3}{c}{$0.115_{-0.034}^{+0.030}$} \\
extra light $\ell_4$ [in \textit{TESS}-band] & \multicolumn{3}{c}{$0.054_{-0.036}^{+0.025}$} & \multicolumn{3}{c}{$0.007_{-0.005}^{+0.009}$} \\
$(M_V)_\mathrm{tot}$  & \multicolumn{3}{c}{$0.03_{-0.09}^{+0.10}$} & \multicolumn{3}{c}{$2.88_{-0.14}^{+0.13}$} \\
distance [pc]         & \multicolumn{3}{c}{$..._{}^{}$} & \multicolumn{3}{c}{$623_{-15}^{+19}$}  \\  
\hline
\end{tabular}

\textit{Notes. } $^a$: $\mathcal{T}_0^\mathrm{inf}$ denotes the moment of an inferior conjunction of the secondary (Ab) and the tertiary (B) along their inner and outer orbits, respectively; $^b$: time of periastron passage; $^c$: for the meaning, significance and discussion of these parameters, see Sect.~\ref{sec:photodynamical}; $^d$: gravitational darkening exponents ($\beta$) and bolometric albedos ($A$) were adjusted only for the inner pair of TIC\,14839347 and, hence, these are given only in this case; $^d$: taken from independent SED analysis with Gaussian priors.
\end{table*}  

\section{Discussion}
\label{sec:discussion}

\subsection{TIC\,14839347}

As was mentioned above, the analysis of this triple system had to be carried out iteratively, and with extra care. We adopted $m_\mathrm{Aa}=2.50\pm0.13\,\mathrm{M}_\odot$ and $T_\mathrm{B}=5000\pm200$\,K for the mass of the primary and effective temperature of the tertiary, from the separate SED analysis. The inner mass ratio was found to be as low as $q_\mathrm{in}=0.25\pm0.02$, which, together with a near Roche-lobe-filling secondary, make this inner binary appear to be a typical Algol-type system with a reversed mass ratio. The distant third star was found to be a little less massive than the current primary of the inner binary ($m_\mathrm{B}=2.36_{-0.09}^{+0.17}\,\mathrm{M}_\odot$).

This finding again emphasizes that, most likely, intensive mass exchange has occurred between the two inner binary stars in the near past of this triple system, as the tertiary component is evidently an evolved, red giant star ($R_\mathrm{B}=8.2\pm0.3\,\mathrm{R}_\odot$), in contrast to the similarly massive, hot ($T_\mathrm{Aa}=8975\pm380$\,K) and much less evolved ($R_\mathrm{Aa}=2.87\pm0.03\,\mathrm{R}_\odot$) primary component.

Here we emphasize again that, in the case of this particular system, we took into account the reflection/irradiation effect and, moreover, both the two gravity-darkening exponents and the bolometric albedos of the inner pair of stars were freely adjusted MCMC parameters. In the case of the hot and, hence, radiative primary star (component Aa) we obtained an albedo of $A_\mathrm{Aa}=0.97\pm0.05$ which fits nicely with the theoretically expected value of $A_\mathrm{Aa}^\mathrm{theo}=1$.\footnote{In this regard, note that in their study \citet{2011MNRAS.414.2413S} have obtained a clearly unphysical value of $A>2$ during the analysis of another semi-detached system with a hot primary. The authors explained this by the incompleteness of the physical model that they used -- which happens to be very close to ours. Hence, this agreement we find with the theoretical value is far from trivial.} On the other hand, the bolometric albedo obtained for the convective secondary ($A_\mathrm{Ab}=0.76\pm0.05$) significantly differs from the theoretically expected value of $A_\mathrm{Ab}^\mathrm{theo}=0.5$. According to our knowledge, such a high albedo for a convective star is quite unusual, however, a deeper investigation of this question is beyond the scope of this paper.

\begin{figure}
\begin{center}
\includegraphics[width=0.99 \columnwidth]{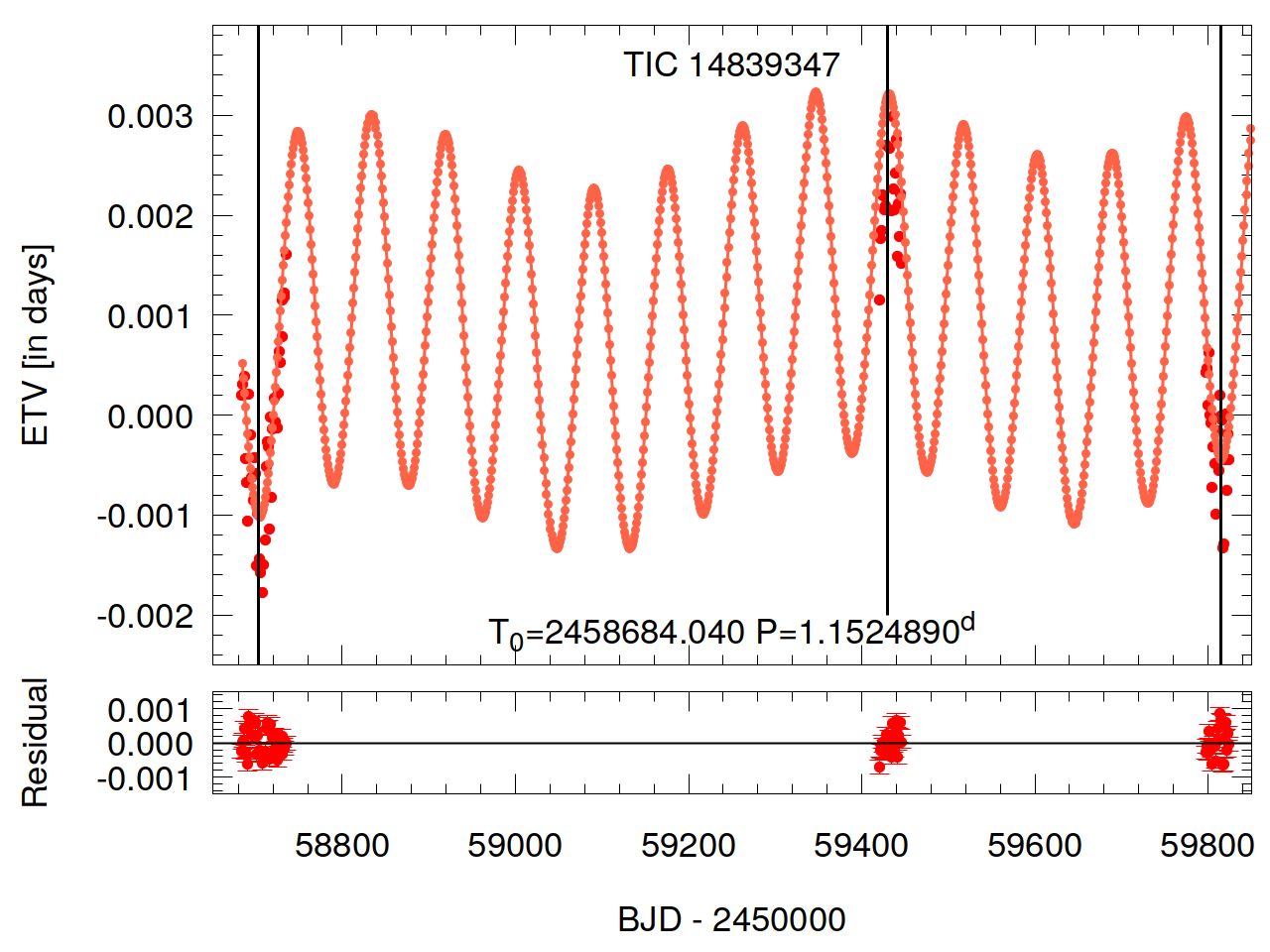}
\caption{Photodynamical fit to the primary \textit{TESS} ETV curve for TIC\,14839347. The larger and darker red circles represent the observed primary times of EB eclipses, while the smaller, lighter symbols, connected with straight lines are taken from the photodynamical model ETV curve. The three thin vertical lines denote the locations of the three third-body outer eclipses. Residuals are also shown in the lower panel, where the uncertainty on each point is also noted.}
\label{fig:T014839347ETV}
\end{center}
\end{figure}

Turning to the orbital properties of this triple system, we find that the inner orbit is circular, as is expected in the case of a nearly semi-detached system.\footnote{Note that though our statistical results give an inner eccentricity of $e_\mathrm{in}=0.0005\pm0.0004$, this is only for the instantaneous osculating orbital elements, which, in general, cannot result in exactly zero eccentricity in a perturbed Keplerian problem. This question was discussed in detail e.g. in \citet{1998MNRAS.300..292K,2002A&A...392..895B}.} By contrast, the outer orbit displays some small, but significant, eccentricity ($e_\mathrm{out}=0.04\pm0.01$). What makes this result, however, a bit less robust is that the secondary outer eclipses (in Sectors 14 and 55) are located exactly at phase 0.5. This is reflected by the fact that the outer argument of pericenter is $\omega_\mathrm{out}=270\degr\pm3\degr$ and, hence, $e_\mathrm{out}\cos\omega_\mathrm{out}\approx0$. In turn, this would suggest that the outer orbit is seen exactly from the direction of its major axis. However, one should keep in mind that, in this particular case, the outer eccentricity is determined only via the weakly determined parameter $e_\mathrm{out}\sin\omega_\mathrm{out}$, which primarily manifests itself in the durations of the outer eclipses and has strong correlations with the outer inclination, the fractional radius of the tertiary, the node of the outer orbit relative to the inner one, and even some other parameters.\footnote{As one can readily see in Fig.~\ref{fig:T014839347ETV}, the ETV curve is currently so poorly covered that it does not carry any information about the eccentricity of the outer orbit.}

As both the inner and outer orbits are fairly circular and almost coplanar ($i_\mathrm{m}=3\fdg5_{-1}^{+3}$), one cannot expect large perturbations and, hence, there are only small departures from Keplerian motions for the inner and outer orbits. This is true despite the fact that the nominal apsidal motion period of the inner orbit is shorter than a year ($P_\mathrm{apse}=0.80\pm0.02$\,yr). As one can see from the different kinds of apsidal advance rates $\Delta\omega$, in the case of the inner orbit the classic tidal ($\Delta\omega_\mathrm{tide}$) contribution is the dominant one---larger by one order of magnitude than that of the third-body driven apsidal motion. We emphasize again, however, that as the inner orbit is practically circular, this very short apsidal motion is only virtual, stemming from the above mentioned slight imperfections in the use of osculating orbital elements in this particular case. On the other hand, for the small but clearly non-zero mutual inclination, one can expect a very low (few degrees) amplitude inclination variation with a period of $P_\mathrm{node}^\mathrm{dyn}=51\pm1$\,yr, which might manifest itself in very small eclipse depth variations in the future.

Here we note also that the SED portion of the iterative solution for this system allowed for an age determination of the system of $520 \pm 115$ Myr\footnote{Here we assumed that even though the inner binary stars may have previously exchanged mass, the tertiary star has not been affected and is evolving normally as a single star.}, and an independent fit for the interstellar extinction of $A_V = 2.2 \pm 0.16$.

\subsection{TIC\,66893949}

\begin{figure}
\begin{center}
\includegraphics[width=0.99 \columnwidth]{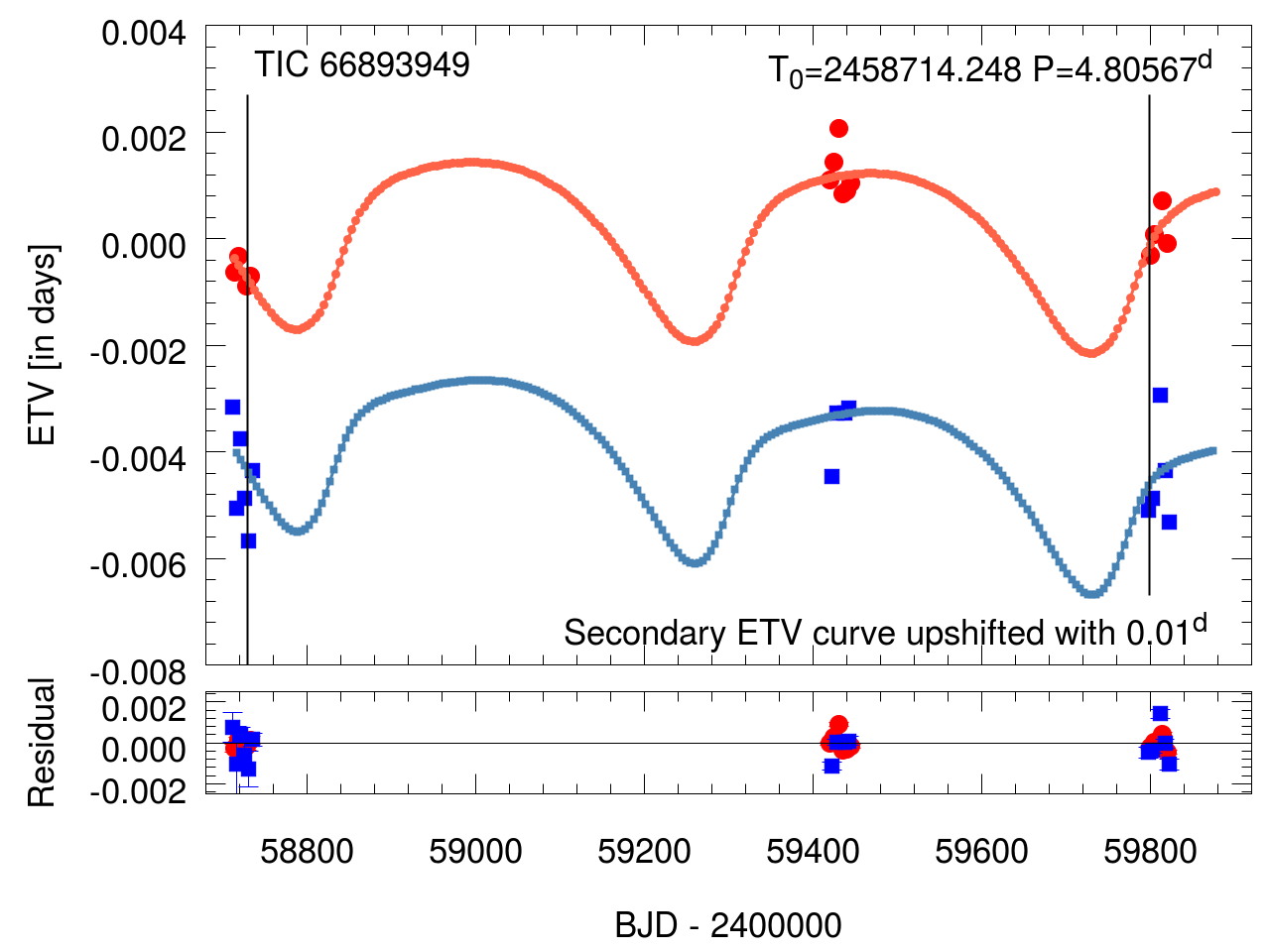}
\caption{Photodynamical fit to the \textit{TESS} ETV curves for TIC \,6893949. The larger and darker red circles and blue squares represent the observed primary and secondary times of EB eclipses, while the smaller, lighter symbols, connected with straight lines are taken from the photodynamical model ETV curve. The two thin vertical lines denote the locations of the two third-body outer eclipses. Residuals are also shown in the lower panel, where the uncertainty on each point is also noted.}
\label{fig:T066893949ETV}
\end{center}
\end{figure}

This system has the longest inner ($P_\mathrm{in}=4.805$\,d) and outer ($P_\mathrm{out}=471.0$\,d) periods. Because the \textit{TESS} observations cover only 16-17\% of the full outer orbit (see Fig.~\ref{fig:T066893949ETV}) it is quite surprising (and, of course, fortuitous) that both primary and secondary third-body eclipses were observed.

According to our photodynamical results, the system consists of three sun-like, main sequence stars ($m_\mathrm{Aa}=1.51\pm0.04\,\mathrm{M}_\odot$, $m_\mathrm{Ab}=1.01\pm0.02\,\mathrm{M}_\odot$, $m_\mathrm{B}=0.97\pm0.04\,\mathrm{M}_\odot$). Note, amongst our four triple systems, this is the only one where the tertiary star was found to be the least massive.

In connection with the somewhat larger inner orbital separation, the inner binary pair has a small, but significant, eccentricity ($e_\mathrm{in}=0.005\pm0.001$). The outer orbit also has the largest eccentricity in our sample with $e_\mathrm{out}=0.402\pm0.004$. Interestingly, despite the somewhat wide configuration, this triple was found to be quite flat ($i_\mathrm{mut}=0\fdg7\pm0\fdg5$).

The dynamical timescales in this systems exceed several centuries for both the inner and outer subsystems. Moreover, as one can see from the separate apsidal advance rates, both the classic tidal and the relativistic contributions to the apsidal motions are negligible relative to the dynamical (forced by third-body perturbations) apsidal motions of the inner and outer orbits.

Finally, we note that our analysis resulted in a slightly larger distance than that of \citet{2021AJ....161..147B}, i.e., $d_\mathrm{phot}=625\pm17$\,pc, vs. $d_\mathrm{EDR3}=587\pm9$\,pc, a $2-\sigma$ discrepancy.

\subsection{TIC\,88206187}

Despite the fact that this triple was observed only in two \textit{TESS} sectors, we were able to obtain the most robust solution for this system among those studied in this work.  The reason is that, fortunately, \textit{TESS} observed both primary and secondary third-body eclipses and, moreover, both the inner and outer eclipses are total (i.e., having flat-bottom mid-eclipse sections) which strongly constrain the surface brightness ratios of the constituent stars.  Moreover, the combination of the \textit{TESS} observations with the archive ground-base data (see Sect.~\ref{Sect:ASASSN-ATLAS}) we were able to determine the outer period with considerable accuracy ($P_\mathrm{out}=52.84\pm0.01$\,days) which was found to be in reasonable agreement with that of the somewhat less accurate \textit{Gaia} SB1 solution ($P_\mathrm{out}=53.03\pm0.06$\,days; Table~\ref{tab:comparing_table}). This makes this triple the most compact in our sample.

\begin{figure}
\begin{center}
\includegraphics[width=0.99 \columnwidth]{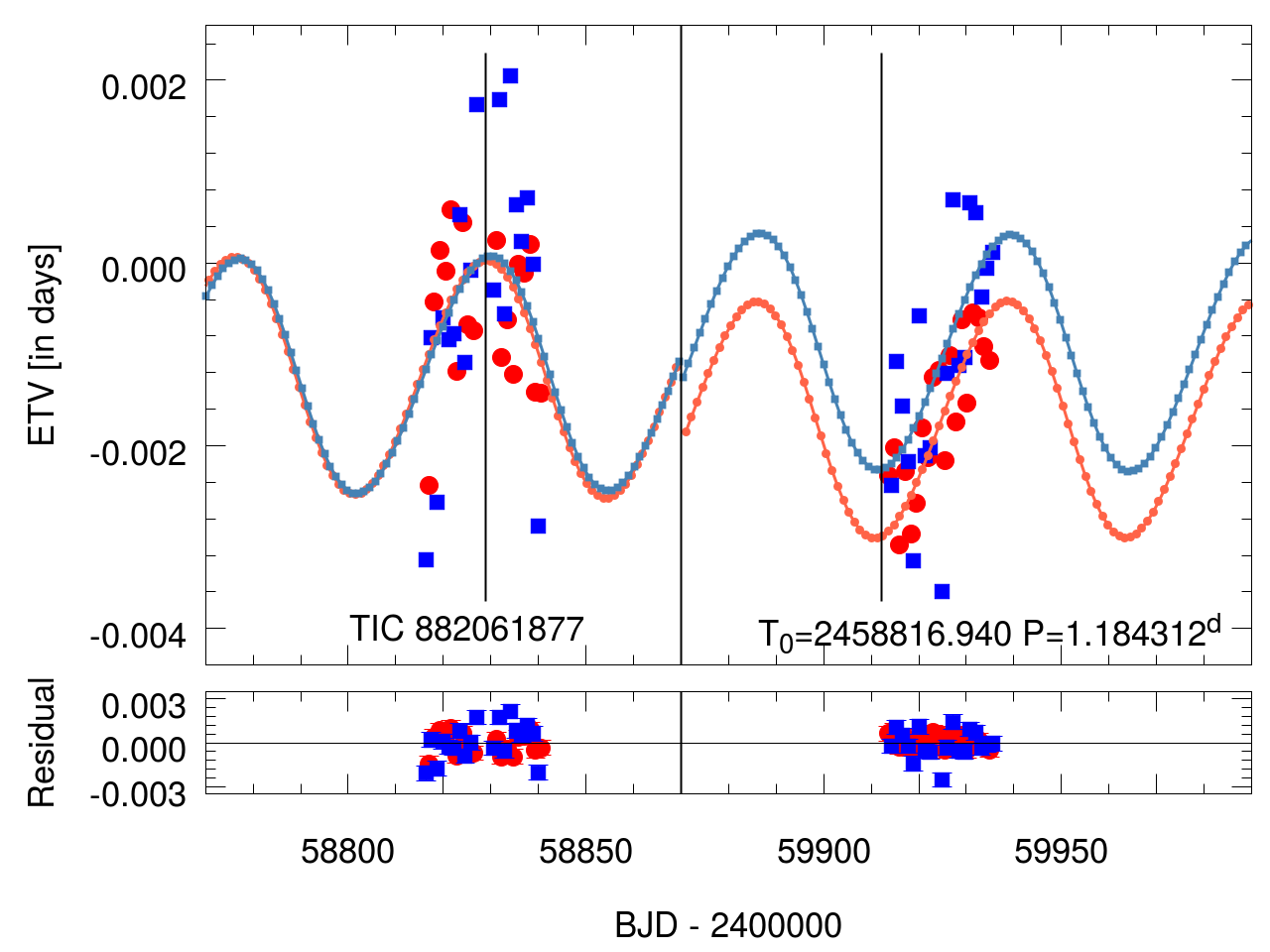}
\caption{Photodynamical fit to the \textit{TESS} ETV curves for TIC\,88206187.}
\label{fig:T088206187ETV}
\end{center}
\end{figure}

Already, a very first inspection of the 2-day-long primary third-body eclipse around BJD~2\,458\,829 (at the middle of Sector 19, see lower left panel of Fig.~\ref{fig:triples}) reveals that the tertiary is likely the largest component and, most probably, is a red giant star.\footnote{Note that amongst other \textit{TESS}-discovered triply eclipsing triples, analyzed in former papers, it is TIC~54060695 whose LC (more strictly speaking, the third-body eclipses) most closely resembles that of the current triple system \citep[see the middle left panel of Fig.~1 of][]{2022MNRAS.513.4341R}.}  The unequal primary and secondary third-body eclipse depths, with the flat bottom of the much deeper (primary) third-body eclipse, indicate that the members of the inner EB are much smaller, but substantially hotter, than the tertiary star.  From this, one can directly infer that the tertiary component should be the most massive amongst the three stars, being the most evolved (assuming, of course, coeval evolution for the three objects). Finally, the fact that, in the case of the primary third-body eclipse (at Sector 19) the tertiary component passes its inferior conjunction point, reveals the outer orbital phase at that (or any other) moment uniquely, despite the absence of any informative ETV data (see Fig.~\ref{fig:T088206187ETV}).

The photodynamical solution then confirmed all these preliminary assessments, as it was found that the tertiary star was actually a moderately massive ($m_\mathrm{B}=2.6\pm0.1\,\mathrm{M}_\odot$) red giant star with the basic parameters of $R_\mathrm{B}=11.7\pm0.3\,\mathrm{R}_\odot$ and $T_\mathrm{B}=4950\pm70$\,K.  The members of the inner binary pair, however, were found to be A and F-type, but still MS stars, with parameters of $m_\mathrm{Aa,Ab}=2.05\pm0.05\,\mathrm{M}_\odot$, $1.35\pm0.03\,\mathrm{M}_\odot$; $R_\mathrm{Aa,Ab}=2.25\pm0.05\,\mathrm{R}_\odot$, $1.35\pm0.03\,\mathrm{R}_\odot$; $T_\mathrm{Aa,Ab}=8\,360^{+270}_{-160}$\,K, $6\,590^{+170}_{-70}$\,K for the primary and secondary binary components, respectively.

The future evolution of this triple seems to be quite interesting. Due to the compactness of both the inner and outer subsystems, we may expect that both the currently more massive red giant tertiary and the TAMS-aged primary of the inner binary will fill out their respective Roche-lobes. The current size of the tertiary and its Roche lobe radius are 11.7\,R$_\odot$ and 38.3\,R$_\odot$, respectively.  The same quantities for the primary EB star, Aa, are 2.25\,R$_\odot$ and 2.94\,R$_\odot$, respectively.  Thus, star Aa needs to increase its radius only by $\simeq 30\%$ in order to overflow its Roche lobe, whereas the tertiary would need to triple its current radius to overflow its Roche lobe.  Nonetheless, according to MIST stellar evolution tracks \citep{dotter16,choi16} it will take star Aa some 260 Myr to reach a state where it overflows its Roche lobe with respect to star Ab.  By contrast, the more massive and currently more evolved tertiary star will evolve much more rapidly and will reach the tip of the red giant branch (RGB) in a few Myr.  At that time, its radius will not quite be sufficiently large to fill its Roche lobe.  However, in less than 180 Myr after that, it will ascend the asymptotic giant branch (AGB) and overfill its Roche lobe for certain. 
 
The detailed evolution of this system, after mass transfer from the giant tertiary to the inner binary commences, is beyond the scope of this paper.  However, we can speculate that since the tertiary will be a convective giant at that time, and the outer mass ratio ($m_B/(m_{\rm Aa}+m_{\rm Ab}$)) is 0.76, the mass transfer could be dynamically stable. But, formally, for a completely convective donor star, this ratio should be $\lesssim 2/3$.

Apart from its future evolution with mass transfer, this triple may harbour other interesting effects in its present configuration as well. Recently \citet{2023MNRAS.521.2114G} called attention to the fact that such compact triple stars, where the tertiaries are red giants, may produce remarkable tidal effects whereby tidal dissipation could lead to observable orbital shrinkage within some decades-long timescales. In this regard, we note that amongst all the known analysed compact triple star systems, TIC~88206187 is only the third case in which the fractional radius of the distant tertiary exceeds 0.1 ($r_\mathrm{B}=R_\mathrm{B}/a_\mathrm{out}=0.109\pm0.002$)\footnote{The other two systems are TIC~242132789 \citep{2022MNRAS.513.4341R} and HD~181068 \citep{2013MNRAS.428.1656B,2013MNRAS.429.2425F} with $r_\mathrm{B}=0.151$ and $0.138$, respectively.}, and thus, tertiary tides might currently be effective, at least marginally. In this regard, also note that, despite the relatively more significant apsidal advance rate of the outer orbit due to tidal forcing ($\Delta\omega_\mathrm{tide}=9''\pm1''$/cycle), it is clearly negligible relative to the third-body effects ($\Delta\omega_\mathrm{tide}=11400''\pm100''$/cycle) and does not alter the timescale of the apsidal precession ($P_\mathrm{apse}=185\pm1$\,yr).

Finally, we note that the photometric distance inferred from our analysis is $d_\mathrm{phot}=2580\pm60$\,pc, which is in quite a good agreement with the \textit{Gaia}-parallax inferred distance of $d_\mathrm{EDR3}=2490\pm90$\,pc \citep{2021AJ....161..147B}.

\subsection{TIC\,298714297}

In contrast to the other systems in this study, TIC~298714297 is found to be a triplet of three low-mass cool red (K and M-type) dwarfs. Besides the three dips of a complex third-body eclipse, the other remarkable feature of the LC is the likely rotational distortions, caused by stellar spots, which are most pronounced in the Sector 55, 56 LCs. Moreover, the system shows three sudden, short $\sim$$3-4\%$ brightenings which we attribute to stellar flares.

All these features suggest strong chromospheric/photospheric stellar activity on at least one of the stellar components.  In the case of the rotational modulation, since the observed period is equal, or very close to, the period of the inner EB, it seems quite likely that its origin is component Aa (which is the most massive and brightest amongst the three stars). In the case of the three stellar flare events, their origin, strictly speaking, is less certain.  But, considering the other signals of strong magnetic activity are likely from star Aa, we may tentatively assume that the flare events are also hosted by this star. It is noteworthy, however, that the two larger flares were observed during Sector 15, when the rotational modulation (i.e., due to the spottedness) was much less pronounced.

Regarding the astrophysical parameters, this triple was found to be not only the least massive ($m_\mathrm{Aa}=0.83\pm0.03\,\mathrm{M}_\odot$, $m_\mathrm{Ab}=0.50\pm0.02\,\mathrm{M}_\odot$, $m_\mathrm{B}=0.68\pm0.02\,\mathrm{M}_\odot$) of our set, but also the most aged ($\tau\sim10$\,Gyr). Note, however, that some caution is necessary, as our model solution resulted in a reddening of $E(B-V)=0.3\pm0.08$ which looks quite unrealistic for such a close object ($d_\mathrm{phot}=138\pm4$\,pc vs. $d_\mathrm{EDR3}=111.9\pm0.7$\,pc \citealt{2021AJ....161..147B}).

\begin{figure}
\begin{center}
\includegraphics[width=0.99 \columnwidth]{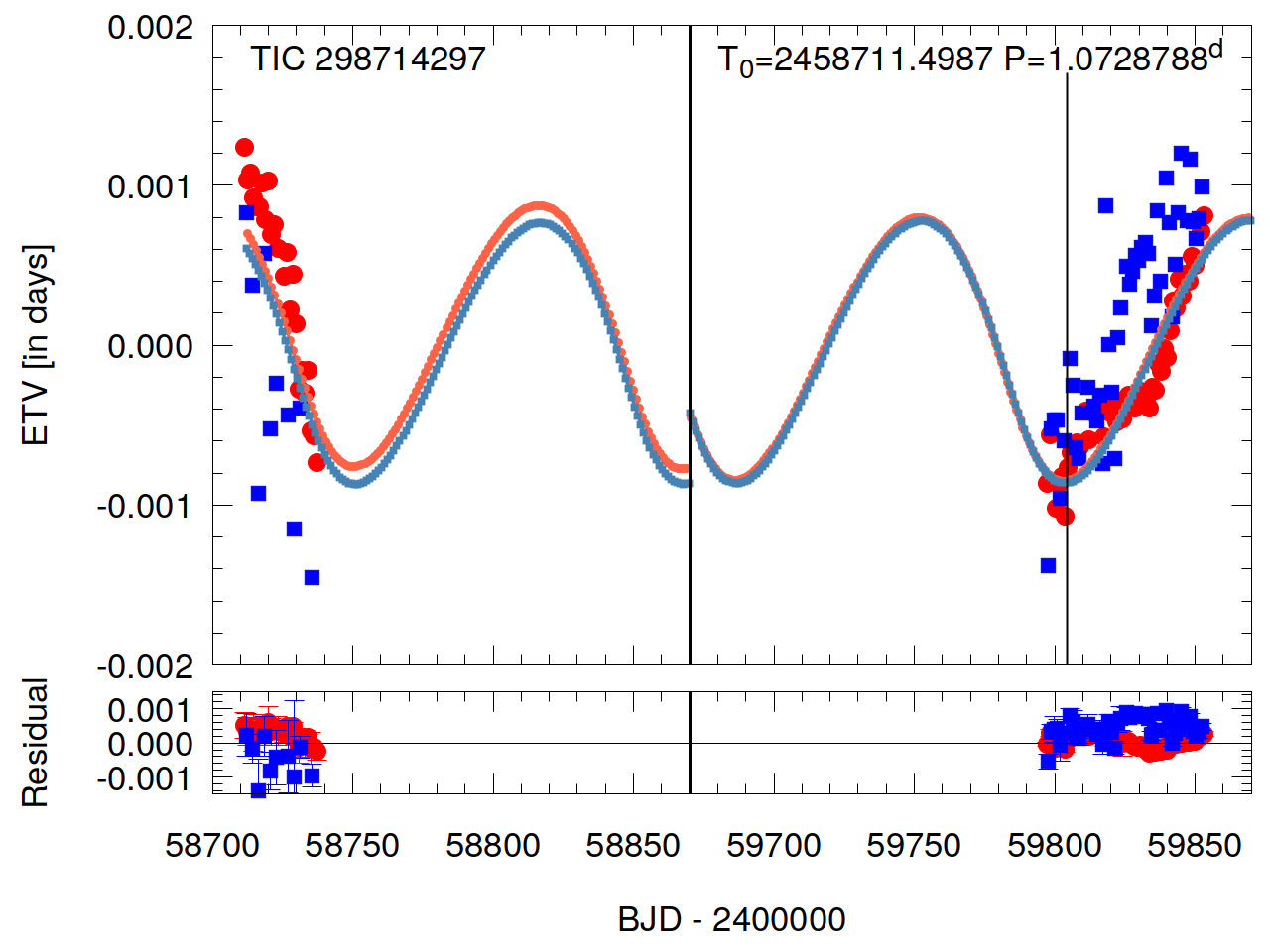}
\caption{Photodynamical fit to the \textit{TESS} ETV curves for TIC\,298714297.}
\label{fig:T298714297ETV}
\end{center}
\end{figure}

The inner orbit was found to be practically circular ($e_\mathrm{in}=0.002_{-0.001}^{+0.007}$) which is naturally expected for such an old and close binary, while the outer orbit was found to be moderately eccentric ($e_\mathrm{out}=0.24\pm0.03$). In this regard, we note that the secondary's ETV curve in the inner EB is offset a little from the primary's one (Fig.~\ref{fig:T298714297ETV}).  However, in our interpretation, in this case such an offset is not due a slight orbital eccentricity, but rather is caused by the LC variations due to the starspots on at least one of the stars.  Such shifts in the ETVs were found by \citet{2013ApJ...774...81T}.

Finally, note also that besides TIC\,14839347, this is the other system in our sample where the apsidal motion of the inner orbit is dominated by the tidal forces instead of third-body perturbations ($\Delta\omega_\mathrm{tide}=105''\pm10''$/cycle vs. $\Delta\omega_\mathrm{3b}=63''\pm2''$/cycle).  But, again, due to the very low inner eccentricity one cannot expect strong, non-Keplerian variations during the apsidal motion cycle of $P_\mathrm{apse}=28\pm2$\,yr.

\begin{table*}
 \centering
\caption{Orbital and astrophysical parameters of TICs 88206187 and 298714297 from the joint photodynamical \textit{TESS}, ETV, SED and \texttt{PARSEC} isochrone solution. Note that the orbital parameters are instantaneous, osculating orbital elements and are given for epoch $t_0$ (first row).  }
 \label{tab:syntheticfit_TIC088206187298714297}
\begin{tabular}{@{}lllllll}
\hline
&\multicolumn{3}{c}{TIC\,88206187} & \multicolumn{3}{c}{TIC\,298714297}  \\
\hline
\multicolumn{7}{c}{orbital elements} \\
\hline
   & \multicolumn{3}{c}{subsystem} & \multicolumn{3}{c}{subsystem} \\
   & \multicolumn{2}{c}{Aa--Ab} & A--B & \multicolumn{2}{c}{Aa--Ab} & A--B \\
  \hline
  $t_0$ [BJD - 2400000] & \multicolumn{3}{c}{58\,816.0} & \multicolumn{3}{c}{58\,711.0} \\
  $P$ [days] & \multicolumn{2}{c}{$1.184592_{-0.000063}^{+0.000055}$} & $52.922_{-0.039}^{+0.041}$ & \multicolumn{2}{c}{$1.072891_{-0.000019}^{+0.000010}$} & $117.24_{-0.31}^{+0.36}$ \\
  $a$ [R$_\odot$] & \multicolumn{2}{c}{$7.085_{-0.061}^{+0.053}$} & $107.7_{-1.0}^{+1.1}$ & \multicolumn{2}{c}{$4.863_{-0.077}^{+0.050}$} & $127.5_{-1.8}^{+1.3}$ \\
  $e$ & \multicolumn{2}{c}{$0.0012_{-0.0005}^{+0.0005}$} & $0.026_{-0.013}^{+0.017}$ & \multicolumn{2}{c}{$0.0019_{-0.0009}^{+0.0072}$} & $0.241_{-0.028}^{+0.029}$ \\
  $\omega$ [deg] & \multicolumn{2}{c}{$75_{-27}^{+21}$} & $147_{-99}^{+100}$ & \multicolumn{2}{c}{$84_{-10}^{+71}$} & $28_{-6}^{+8}$ \\ 
  $i$ [deg] & \multicolumn{2}{c}{$89.86_{-1.98}^{+2.36}$} & $89.54_{-0.75}^{+0.73}$ & \multicolumn{2}{c}{$89.00_{-0.79}^{+0.93}$} & $89.86_{-0.05}^{+0.10}$  \\
  $\mathcal{T}_0^\mathrm{inf/sup}$ [BJD - 2400000] & \multicolumn{2}{c}{$58\,816.9391_{-0.0002}^{+0.0002}$} & $58\,828.8819_{-0.0081}^{+0.0081}$ & \multicolumn{2}{c}{$58\,711.4972_{-0.0001}^{+0.0002}$} & ${59\,804.6649_{-0.0385}^{+0.0396}}^*$  \\
  $\tau$ [BJD - 2400000] & \multicolumn{2}{c}{$58\,816.292_{-0.094}^{+0.067}$} & $58\,800.2_{-13.8}^{+15.1}$ & \multicolumn{2}{c}{$58\,710.942_{-0.029}^{+0.213}$} & $59\,791.8_{-1.6}^{+2.0}$  \\
  $\Omega$ [deg] & \multicolumn{2}{c}{$0.0$} & $-1.17_{-2.18}^{+3.01}$ & \multicolumn{2}{c}{$0.0$} & $4.23_{-4.13}^{+4.33}$ \\
  $i_\mathrm{mut}$ [deg] & \multicolumn{3}{c}{$2.90_{-1.37}^{+1.63}$} & \multicolumn{3}{c}{$4.41_{-2.70}^{+4.18}$} \\
  \hline
  mass ratio $[q=m_\mathrm{sec}/m_\mathrm{pri}]$ & \multicolumn{2}{c}{$0.661_{-0.009}^{+0.009}$} & $0.765_{-0.017}^{+0.010}$ & \multicolumn{2}{c}{$0.617_{-0.016}^{+0.013}$} & $0.512_{-0.009}^{+0.007}$  \\
$K_\mathrm{pri}$ [km\,s$^{-1}$] & \multicolumn{2}{c}{$120.36_{-0.92}^{+1.02}$} & $44.48_{-0.73}^{+0.97}$ & \multicolumn{2}{c}{$87.26_{-2.02}^{+2.14}$} & $19.12_{-0.27}^{+0.34}$ \\ 
$K_\mathrm{sec}$ [km\,s$^{-1}$] & \multicolumn{2}{c}{$182.26_{-2.40}^{+1.83}$} & $58.47_{-0.37}^{+0.41}$ & \multicolumn{2}{c}{$141.79_{-1.94}^{+1.14}$} & $37.48_{-0.60}^{+0.57}$ \\ 
  \hline  
\multicolumn{7}{c}{Apsidal and nodal motion related parameters} \\
\hline  
$P_\mathrm{apse}$ [year] & \multicolumn{2}{c}{$7.3_{-0.3}^{+0.2}$} & $185_{-1}^{+1}$  & \multicolumn{2}{c}{$27.8_{-2.2}^{+1.9}$} & $1124_{-34}^{+30}$ \\
$P_\mathrm{apse}^\mathrm{dyn}$ [year] & \multicolumn{2}{c}{$5.2_{-0.2}^{+0.1}$} & $16.41_{-0.09}^{+0.16}$ & \multicolumn{2}{c}{$22.4_{-1.5}^{+1.2}$} & $104_{-3}^{+2}$ \\
$P_\mathrm{node}^\mathrm{dyn}$ [year] & \multicolumn{3}{c}{$18.0_{-0.2}^{+0.1}$} & \multicolumn{3}{c}{$114_{-2}^{+3}$} \\
$\Delta\omega_\mathrm{3b}$ [arcsec/cycle] & \multicolumn{2}{c}{$443_{-5}^{+4}$} & $11433_{-113}^{+62}$ & \multicolumn{2}{c}{$63.4_{-1.5}^{+1.6}$} & $4013_{-85}^{+105}$ \\
$\Delta\omega_\mathrm{GR}$ [arcsec/cycle] & \multicolumn{2}{c}{$3.96_{-0.07}^{+0.06}$} & $0.459_{-0.008}^{+0.10}$ & \multicolumn{2}{c}{$2.28_{-0.07}^{+0.05}$} & $0.014_{-0.004}^{+0.004}$ \\
$\Delta\omega_\mathrm{tide}$ [arcsec/cycle] & \multicolumn{2}{c}{$362_{-17}^{+24}$} & $9.0_{-0.8}^{+0.9}$ & \multicolumn{2}{c}{$104_{-9}^{+12}$} & $0.025_{-0.002}^{+0.002}$ \\
\hline
\multicolumn{7}{c}{stellar parameters} \\
\hline
   & Aa & Ab &  B & Aa & Ab &  B \\
  \hline
 \multicolumn{7}{c}{Relative quantities} \\
  \hline
 fractional radius [$R/a$] & $0.3185_{-0.0039}^{+0.0047}$ & $0.1905_{-0.0028}^{+0.0030}$ & $0.1086_{-0.0020}^{+0.0021}$ & $0.1728_{-0.0025}^{+0.0026}$ & $0.1042_{-0.0043}^{+0.0050}$ & $0.0053_{-0.0001}^{+0.0001}$ \\
 temperature relative to $(T_\mathrm{eff})_\mathrm{Aa}$ & $1$ & $0.789_{-0.010}^{+0.010}$ & $0.592_{-0.012}^{+0.010}$ & $1$ & $0.678_{-0.007}^{+0.006}$ & $0.840_{-0.013}^{+0.015}$ \\
 fractional flux [in \textit{TESS}-band] & $0.1600_{-0.0026}^{+0.0027}$ & $0.0294_{-0.0011}^{+0.0011}$ & $0.800_{-0.015}^{+0.008}$ & $0.6794_{-0.0335}^{+0.0386}$ & $0.0532_{-0.0026}^{+0.0029}$ & $0.2062_{-0.0217}^{+0.0191}$ \\
 \hline
 \multicolumn{7}{c}{Physical Quantities} \\
  \hline 
 $m$ [M$_\odot$] & $2.047_{-0.061}^{+0.051}$ & $1.349_{-0.026}^{+0.031}$ & $2.589_{-0.090}^{+0.107}$ & $0.827_{-0.036}^{+0.024}$ & $0.508_{-0.025}^{+0.024}$ & $0.678_{-0.022}^{+0.026}$ \\
 $R$ [R$_\odot$] & $2.253_{-0.034}^{+0.048}$ & $1.349_{-0.028}^{+0.030}$ & $11.71_{-0.30}^{+0.28}$ & $0.839_{-0.023}^{+0.021}$ & $0.506_{-0.028}^{+0.030}$ & $0.669_{-0.018}^{+0.022}$  \\
 $T_\mathrm{eff}$ [K] & $8358_{-154}^{+279}$ & $6587_{-72}^{+169}$ & $4947_{-59}^{+70}$ & $5332_{-198}^{+106}$ & $3620_{-121}^{+55}$ & $4463_{-137}^{+133}$  \\
 $L_\mathrm{bol}$ [L$_\odot$] & $22.6_{-2.0}^{+2.5}$ & $3.13_{-0.23}^{+0.28}$ & $74.0_{-4.3}^{+4.6}$ & $0.504_{-0.040}^{+0.037}$ & $0.039_{-0.004}^{+0.004}$ & $0.159_{-0.016}^{+0.019}$ \\
 $M_\mathrm{bol}$ & $1.39_{-0.12}^{+0.10}$ & $3.53_{-0.09}^{+0.08}$ & $0.10_{-0.07}^{+0.07}$ & $5.51_{-0.08}^{+0.09}$ & $8.30_{-0.11}^{+0.11}$ & $6.77_{-0.12}^{+0.11}$ \\
 $M_V           $ & $1.35_{-0.08}^{+0.09}$ & $3.50_{-0.09}^{+0.09}$ & $0.38_{-0.09}^{+0.07}$ & $5.67_{-0.10}^{+0.15}$ & $9.82_{-0.19}^{+0.33}$ & $7.41_{-0.23}^{+0.23}$ \\
 $\log g$ [dex] & $4.040_{-0.011}^{+0.010}$ & $4.307_{-0.010}^{+0.010}$ & $2.713_{-0.016}^{+0.016}$ & $4.505_{-0.012}^{+0.010}$ & $4.733_{-0.030}^{+0.027}$ & $4.617_{-0.012}^{+0.009}$ \\
 \hline
\multicolumn{7}{c}{Global system parameters} \\
  \hline
$\log$(age) [dex] & \multicolumn{3}{c}{$8.81_{-0.05}^{+0.04}$} & \multicolumn{3}{c}{$10.02_{-0.04}^{+0.04}$} \\
$[M/H]$  [dex]    & \multicolumn{3}{c}{$0.17_{-0.16}^{+0.07}$} & \multicolumn{3}{c}{$-0.01_{-0.18}^{+0.28}$} \\
$E(B-V)$ [mag]    & \multicolumn{3}{c}{$0.28_{-0.02}^{+0.04}$} & \multicolumn{3}{c}{$0.30_{-0.09}^{+0.05}$} \\
extra light $\ell_4$ [in \textit{TESS}-band] & \multicolumn{3}{c}{$0.010_{-0.008}^{+0.015}$} & \multicolumn{3}{c}{$0.05_{-0.04}^{+0.05}$} \\
$(M_V)_\mathrm{tot}$  & \multicolumn{3}{c}{$-0.03_{-0.08}^{+0.07}$} & \multicolumn{3}{c}{$5.45_{-0.12}^{+0.17}$} \\
distance [pc]         & \multicolumn{3}{c}{$2584_{-68}^{+53}$} & \multicolumn{3}{c}{$138_{-4}^{+4}$}  \\  
\hline
\end{tabular}

\textit{Notes. }{$\mathcal{T}_0^\mathrm{inf/sup}$ denotes the moment of an inferior or superior conjunction of the secondary (Ab) and the tertiary (B) along their inner and outer orbits, respectively. Superior conjunctions are noted with $^*$.}
\end{table*}  

\section{Comparing orbital parameters with \textit{Gaia}}
\label{sect:comparison}

In general, we find substantial agreement between the \textit{Gaia} and photodynamical results for the four systems studied in this work. As far as we can judge, however, our results are generally the more accurate ones. Having said this, however, there are a number of caveats to discuss.  First, we note that a comparison of the orbital elements from the \textit{Gaia} NSS solution and those from the {\sc Lightcurvefactory} photodynamical models is limited due to the differences in their methodologies. The \textit{Gaia} NSS solutions adopt a Keplerian two-body orbit model, whereas the calculated orbital elements from {\sc Lightcurvefactory} represent instantaneous osculating elements. Consequently, the orbital elements obtained from these two approaches are not exactly comparable. However, for the four sources studied in this work, the perturbations to the motions along the outer orbits are negligible on the timescale of the \textit{Gaia} observation interval, and therefore they do not result in significant discrepancies on this account. For example, the apsidal motion timescales of the outer orbits in all our four triples range from 185\,yr to 2820\,yr.  Hence, the orientations of the astronomical orbits measured by \textit{Gaia} will vary by only a negligible amount during the full operation of the space telescope. Moreover, due to the flatness of the four systems, no significant variations in the outer inclinations will occur. One can therefore expect that the orbits will remain (almost) pure Keplerian over the entire life cycle of \textit{Gaia}.

\begin{figure*}
\begin{center}
\includegraphics[width=0.98 \columnwidth]{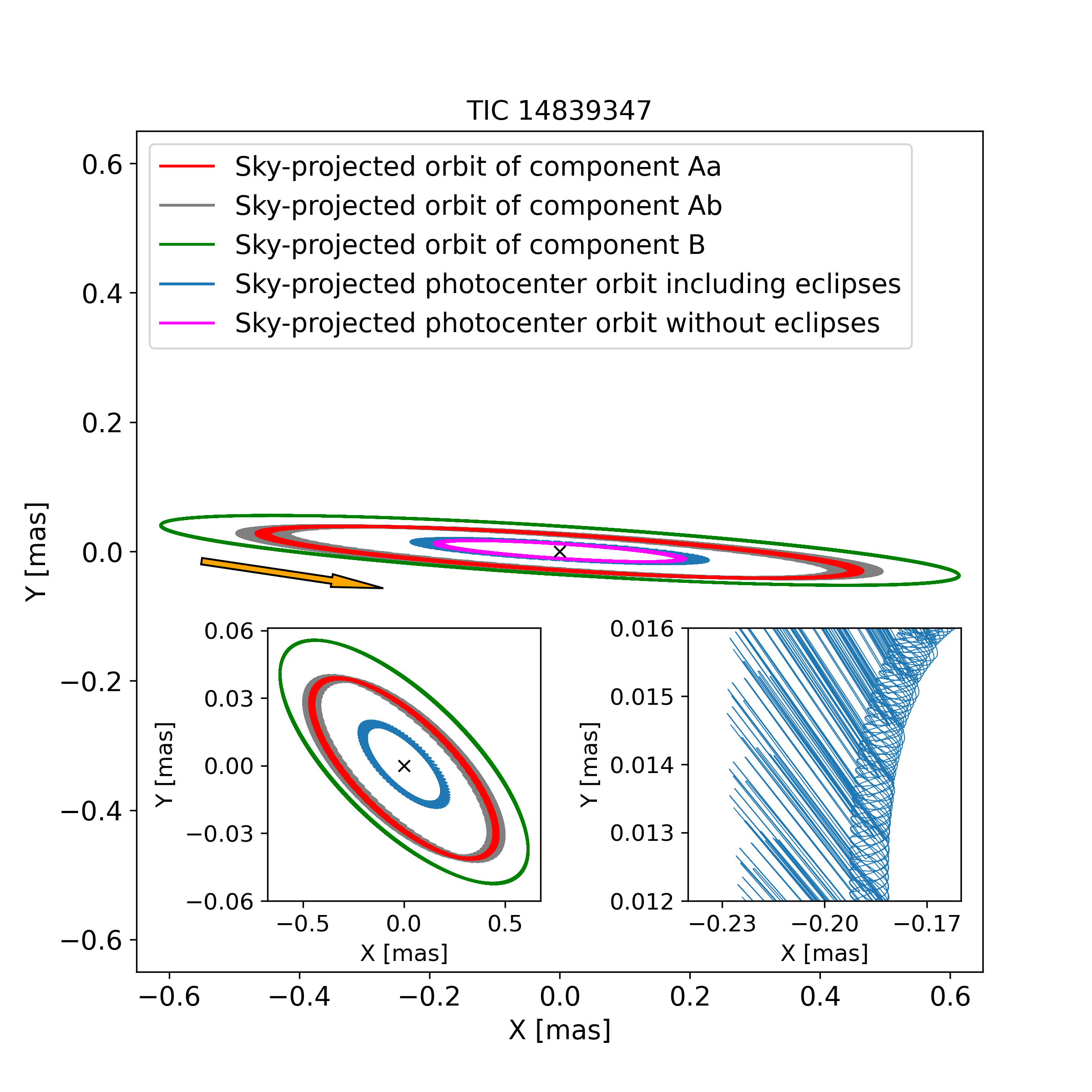}\includegraphics[width=0.98 \columnwidth]{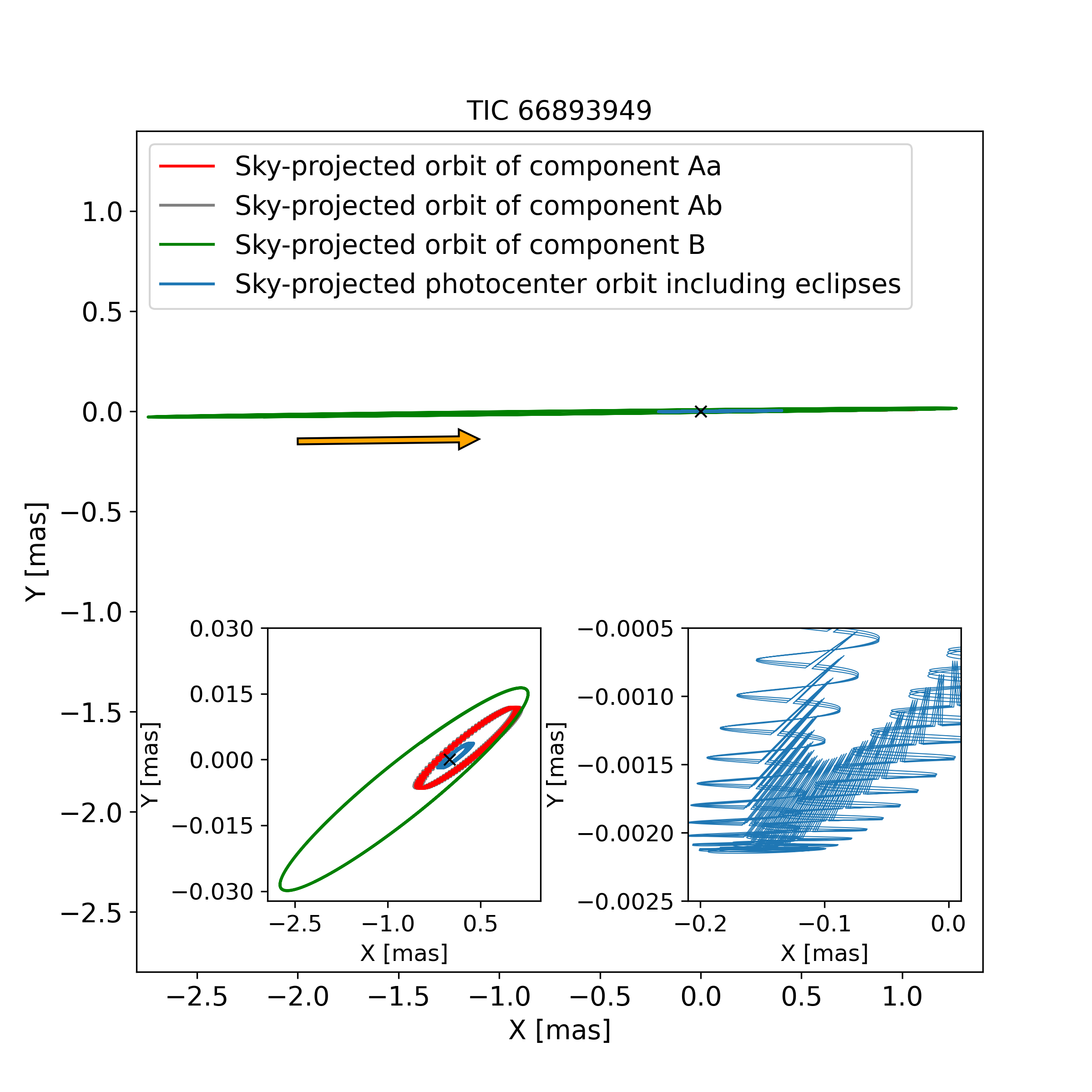}
\includegraphics[width=0.98 \columnwidth]{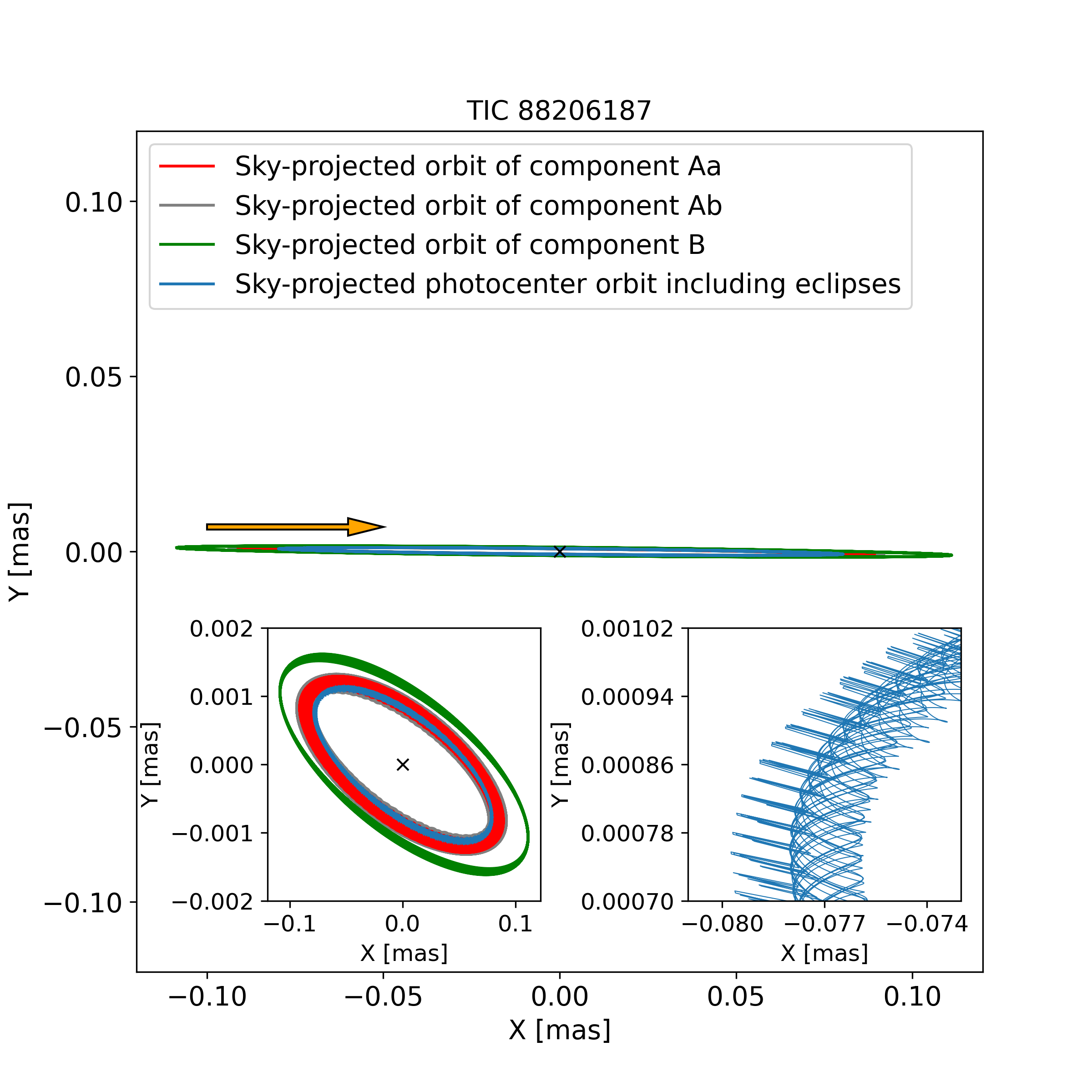}\includegraphics[width=0.98 \columnwidth]{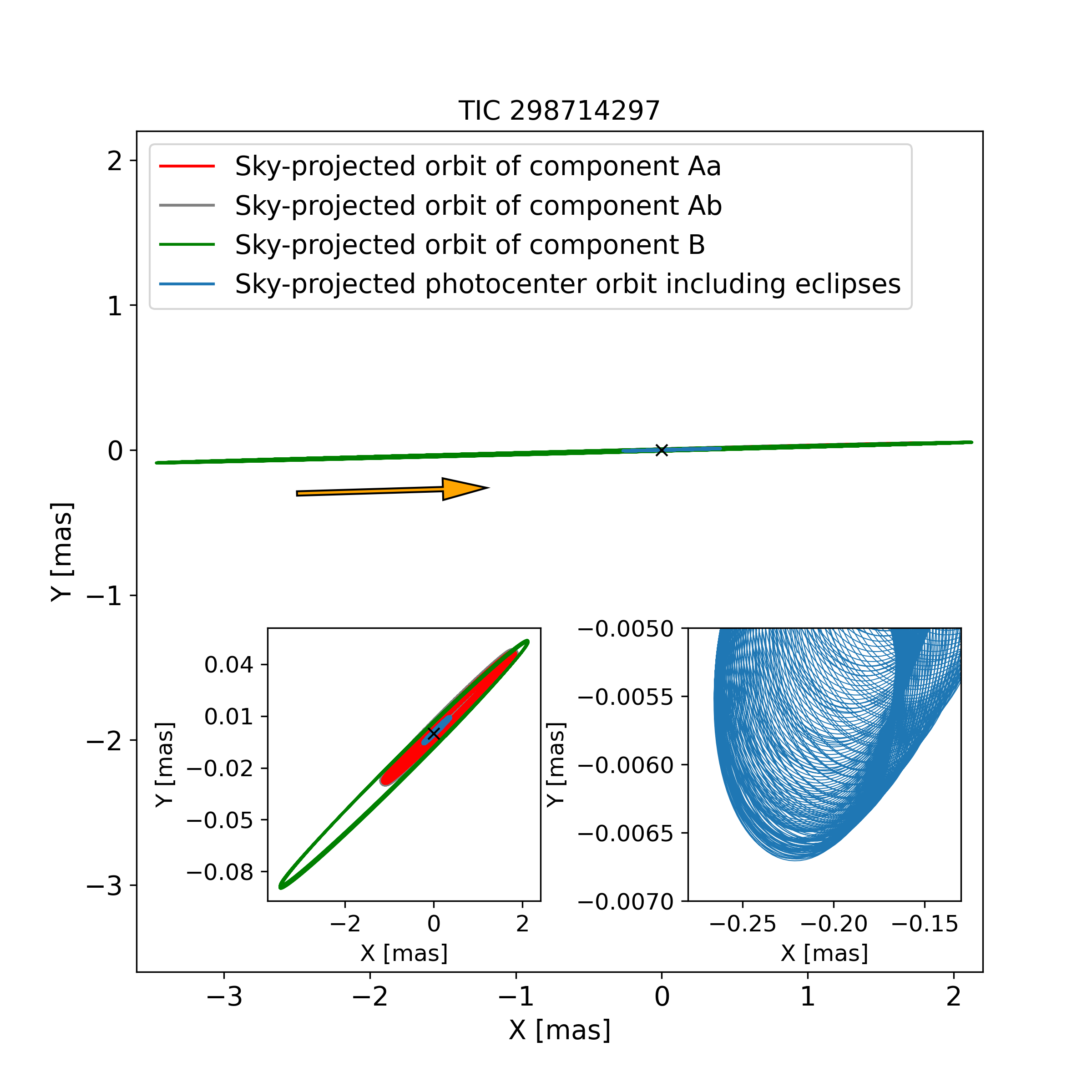}
\caption{Astrometric orbits for each stellar component in all four investigated triple systems, together with our {\it calculated} photocenter orbits, the latter of which are only measurable by \textit{Gaia}. The sky positions and therefore, the astrometric orbits of all the components were taken directly from the numerical integrator of {\sc Lightcurvefactory}.  These were then converted into milliarcseconds from their original physical dimensions according to the photometric distances obtained from the photodynamical solutions. The positions of the photocenters were calculated from the flux-weighted averages of the individual stellar positions, according to Eqs.~(C1--C2) of \citet{2022MNRAS.513.4341R}. For each plot there are also two smaller, insert figures. In the left inserts we show again the astrometric orbits, but with unequally scaled $x$ and $y$ axes for a better view of the nearly edge-on and, consequently, near straight line-like sky-projected orbits. In the right inserts we display zoomed short sections of the photocenter orbits, where the rapid flux variations caused by the inner eclipses are also taken into account. As one can see, these eclipses produce sudden, quasi discontinuous shifts in the photocenter, which may result in higher scatter and, hence, increased uncertainties in the astrometric orbital elements.}
\label{fig:photocenterorbit}
\end{center}
\end{figure*} 

Another issue is that in the \textit{Gaia} DR3 catalogue, the NSS solutions are limited to spatially unresolved stellar systems, primarily due to the specific data processing constraints employed by the \textit{Gaia} team. In the case of astrometric solutions (i.e., for TIC\,298714297 and TIC\,66893949) \textit{Gaia} measures only the orbit of the photocenter of the three stars, and not of any of the individual constituent members within the triple system (\citealt{marcussen23}; Eq. C2 of \citealt{2022MNRAS.513.4341R}). This fact, however, may produce some departures from pure Keplerian astrometric orbits, as during the relatively deep primary eclipses of the four inner pairs, one may expect some quasi-discontinuous shifts in the positions of the photocenter.\footnote{Naturally, the same holds for the astrometric measurements  during third-body eclipses, where these discontinuities might be even much larger. However, due the rarity of the extra eclipse events, observations during these events are less likely.} In order to illustrate these effects, we plot the theoretically expected photocenter orbits (together with the above mentioned potential discontinuities in the small, inserted plots) for all four investigated systems in Fig.~\ref{fig:photocenterorbit}. Keeping in mind all of the above discussion, we conclude that the only direct comparisons that can be made for these systems with the photodynamical results are the period, the eccentricity, and the inclination of the outer orbit.  In the cases involving the \textit{Gaia} spectroscopic orbits (i.e., TIC\,88206187 and TIC\,14839347), \textit{Gaia} primarily observes the spectral lines of the more luminous tertiary component.  Thus we can also compare the \textit{Gaia} RV semi-amplitude ($K1$) of the outer tetriary with the value from the photodynamical fit.  

A third issue is that generally the argument of periastron of the outer orbit from the \textit{Gaia} measurements and those of the photodyamical results typically differ by 180$^\circ$.  This is due to the fact that \textit{Gaia} is generally tracking the brighter tertiary, while the photodynamical analysis utilizes the ETV points from the inner binary.

Finally, with these caveats in mind, we compare the \textit{Gaia} and photodynamical analysis in Table \ref{tab:comparing_table}.  We find generally good agreement for the outer orbital period and its inclination angle.  However, in the case of the outer eccentricity, the differences are mostly larger, but in the case of TIC\,66893949, for example, the two values are in good agreement.


\begin{table*}
\caption{Comparison of orbital parameters of the outer orbit between {\sc LightcurveFactory} and {\it Gaia} DR3 NSS solution. For each object, the upper values belong to the {\sc LightcurveFactory}, while the lower values correspond to the {\it Gaia} DR3 NSS solution.}
\label{tab:comparing_table}
\centering
\begin{tabular}{c | c | c | c | c | c | c | c}
Name & NSS model & P[days] & e & $\mathrm{\omega [deg]}$ & $\mathrm{a [R_{\odot}]}$ & $\mathrm{i [deg]}$ &  $\mathrm{K_{sec} [km/s]}$ \\
\hline
TIC\,14839347 & SB1 & \begin{tabular}{@{}c@{}} $85.524_{-0.0015}^{+0.0017}$ \\ $85.315\pm0.2458$  \end{tabular}& \begin{tabular}{@{}c@{}} $0.047_{-0.013}^{+0.011}$ \\ $0.022\pm0.057$  \end{tabular}& \begin{tabular}{@{}c@{}} $269.9_{-3.3}^{+2.8}$\\ $44.0\pm156.3$  \end{tabular}& - & - & \begin{tabular}{@{}c@{}} $46.4_{-0.7}^{+0.5}$ \\ $50.3\pm2.6$  \end{tabular}\\
\hline
TIC\,66893949 & Orbital &\begin{tabular}{@{}c@{}}$471.03_{-0.07}^{+0.10}$ \\ $470.72\pm3.37$  \end{tabular}&\begin{tabular}{@{}c@{}}$0.4016_{-0.0038}^{+0.0039}$ \\ $0.404\pm0.065$ \end{tabular}&\begin{tabular}{@{}c@{}}$25.8_{-1.5}^{+1.3}$ \\ $237.011\pm 8.744$  \end{tabular}&\begin{tabular}{@{}c@{}}  $386.4_{-3.4}^{+4.2}$ \\ $72.09\pm2.48$  \end{tabular} &\begin{tabular}{@{}c@{}}$90.222_{-0.009}^{+0.010}$ \\$91.39\pm1.37$ \end{tabular}& - \\
\hline
TIC\,88206187 & SB1 &\begin{tabular}{@{}c@{}} $52.922_{-0.039}^{+0.041}$ \\ $53.035\pm0.058$  \end{tabular} &\begin{tabular}{@{}c@{}} $0.026_{-0.013}^{+0.017}$ \\  $0.100\pm0.044$  \end{tabular} &\begin{tabular}{@{}c@{}}$147_{-99}^{+100}$ \\ $212\pm9$   \end{tabular} & - & - & \begin{tabular}{@{}c@{}}$58.47_{-0.37}^{+0.41}$ \\  $65.23\pm3.54$ \end{tabular} \\
\hline
TIC\,298714297 & Orbital & \begin{tabular}{@{}c@{}}  $117.25_{-0.31}^{+0.36}$  \\ $118.60\pm0.28$  \end{tabular} & \begin{tabular}{@{}c@{}}$0.242_{-0.029}^{+0.029}$\\ $0.185\pm0.065$  \end{tabular} & \begin{tabular}{@{}c@{}} $28_{-7}^{+8}$ \\$161\pm20$  \end{tabular} & \begin{tabular}{@{}c@{}}$127.3_{-1.8}^{+1.3}$ \\ $16.37\pm0.50$ \end{tabular} & \begin{tabular}{@{}c@{}} $89.86_{-0.05}^{+0.09}$  \\ $89.53\pm2.21$  \end{tabular} & - \\

\end{tabular}
\end{table*}

\section{Summary and Conclusions}
\label{sect:conclusions}

The most common method for discovering triply eclipsing triple star systems involves searching for evidence of a third-body eclipse in the extended time series of EB systems. These utilize the extensive time series datasets obtained from the \textit{TESS} or \textit{Kepler} photometric sky survey telescopes. The searches are usually done by eye or with the assistance of machine learning tools \citep[see, e.g.,][]{2022MNRAS.513.4341R,2023MNRAS.521..558R}.  It goes without saying that the longer the duration of a data set the greater the likelihood of identifying eclipses involving third bodies within a given system.

In our previous paper \citep{czavalinga23}, we reported the discovery of four triply eclipsing triple systems using an alternative method that utilizes the \textit{Gaia} DR3 data set. In that study, we crossmatched a catalog of approximately 1 million binaries with \textit{Gaia} stars that had longer-period NSS orbital solutions.  Of the $\sim$400 triple systems found in that study, four of them turned out to also be triply eclipsing, as discovered through the use of {\it TESS} data.  Though the fractional yield was small (1\%), this still demonstrated the effectiveness of alternative methods in the search for triply eclipsing triples.

In this work, we studied these four triply eclipsing systems in detail using precision ETV curves and photometry from {\it TESS} data, archival photometry from ASAS-SN and ATLAS, and archival SED curves.  These input data sets undergo a comprehensive analysis via a spectro-photodynamics code ({\sc LightcurveFactory}) to yield both the stellar and orbital parameters of the systems.  In general, the four new compact triples are not dissimilar to ones our group has been studying over the past number of years.

TIC\,66893949 consists of stars with masses similar to our Sun. This system has the longest outer period among the triples investigated in our sample, and it has the most eccentric outer orbit $e_{\rm out} \simeq 0.4$. Even though it has a wide configuration, the mutual inclination angle is fairly low at $i_\mathrm{mut} = 0.6 \pm 0.5$ degrees.

 TIC\,88206187 has the shortest outer period in our sample $P_\mathrm{out} = 52.92 \pm 0.04$ days, and contains a tertiary red giant star with radius of  $11.7 \pm 0.3 \mathrm{R}_{\odot}$. This represents only the third system discovered thus far in which the fractional radius (i.e., $R/a$) of the distant tertiary component exceeds 0.1, suggesting the presence of ongoing tidal effects. Our photodynamical model benefitted from the ATLAS and ASAS-SN LCs which show both the outer primary and secondary eclipses, and therefore provide a precise period and a value for $e_{\rm out} \cos \omega_{\rm out}$. 
 From the MIST stellar evolution tracks, we find that it is likely that the outer tetriary will be the first star in the system to fill its Roche lobe when it becomes an AGB star and transfer mass to the inner binary pair.

TIC\,298714297 is one of the closest triply eclipsing triple systems ever found with a distance of $138\pm4$ pc. It contains low-mass stars that exhibit pronounced spot activity in one of the components of the inner binary, with several flares identified. The most probable host of these flares is star Aa with the largest luminosity in the system.

We encountered some difficulties with modelling TIC~14839347. The low mass secondary component of the inner pair was found to be strongly non-spherical, filling or nearly filling its Roche lobe, which hints at a previous episode of mass transfer. Consequently, the use of the \texttt{PARSEC} single-star evolutionary tracks within the {\sc Lightcurvefactory} analysis could not be used.  However, we were able to use a separate SED fitting code \citep{2022MNRAS.513.4341R} that does not assume any physical relation between the radius and $T_{\rm eff}$ of either binary star, and which can estimate the stellar properties of all three stars. That code, in turn, used estimates of a number of dimensionless ratios such as of the radii and $T_{\rm eff}$ which we could get from {\sc Lightcurvefactory}.  Since neither code could proceed without results from the other, we ran them iteratively until a satisfactory solution was achieved.
From this analysis we find that the inner binary has a mass ratio  of $0.25\pm0.04$ and, as mentioned above, the secondary is nearly filling its Roche lobe. Therefore the inner binary appears as an Algol system with an inverted mass ratio. The outer stellar component is an evolved red giant star, while the primary star of the inner binary appears to be less evolved than the tertiary though these two stars are similar in mass.  This  feature may also be explained by the fact that a mass exchange event has occurred between the inner binary stars in the near past.

The use of \textit{Gaia} DR3 NSS solutions have demonstrated the value of exploring triple systems using alternative methodologies. The upcoming release of \textit{Gaia} DR4 will provide even more extensive data, making it extremely worthwhile to search for signs of triples within these datasets and subsequently perform photodynamical analyses on them.  Such a greatly enhanced statistical sample of compact triples should yield a more comprehensive understanding of these important, intriguing, and fun systems.

\section*{Acknowledgements}
A.\,P. acknowledges the financial support of the Hungarian National Research, Development and Innovation Office -- NKFIH Grant K-138962.

This paper includes data collected by the \textit{TESS} mission. Funding for the \textit{TESS} mission is provided by the NASA Science Mission directorate. Some of the data presented in this paper were obtained from the Mikulski Archive for Space Telescopes (MAST). STScI is operated by the Association of Universities for Research in Astronomy, Inc., under NASA contract NAS5-26555. Support for MAST for non-HST data is provided by the NASA Office of Space Science via grant NNX09AF08G and by other grants and contracts.

This work has made use  of data  from the European  Space Agency (ESA)  mission {\it Gaia}\footnote{\url{https://www.cosmos.esa.int/gaia}},  processed  by  the {\it   Gaia}   Data   Processing   and  Analysis   Consortium   (DPAC)\footnote{\url{https://www.cosmos.esa.int/web/gaia/dpac/consortium}}.  Funding for the DPAC  has been provided  by national  institutions, in  particular the institutions participating in the {\it Gaia} Multilateral Agreement.

We have made use of the All-Sky Automated Survey for Supernovae archival photometric data. See  \citep{shappee14} and \citep{kochanek17} for details of the ASAS-SN survey. We also acknowledge use of the photometric archival data from the Asteroid Terrestrial-impact Last Alert System (ATLAS) project. See \citep{tonry18} and \citep{heinze18} for specifics of the ATLAS survey. 

This publication makes use of data products from the Wide-field Infrared Survey Explorer, which is a joint project of the University of California, Los Angeles, and the Jet Propulsion Laboratory/California Institute of Technology, funded by the National Aeronautics and Space Administration. 

This publication makes use of data products from the Two Micron All Sky Survey, which is a joint project of the University of Massachusetts and the Infrared Processing and Analysis Center/California Institute of Technology, funded by the National Aeronautics and Space Administration and the National Science Foundation.

We  used the  Simbad  service  operated by  the  Centre des  Donn\'ees Stellaires (Strasbourg,  France) and the ESO  Science Archive Facility services (data  obtained under request number 396301). 

This research has made use of the VizieR catalogue access tool, CDS, Strasbourg, France (DOI : 10.26093/cds/vizier). The original description  of the VizieR service was published in \citep{vizier00}.
\section*{Data Availability}

The \textit{TESS} data underlying this article were accessed from MAST (Barbara A. Mikulski Archive for Space Telescopes) Portal (\url{https://mast.stsci.edu/portal/Mashup/Clients/Mast/Portal.html}), including the data products found in the bulk download website (\url{http://archive.stsci.edu/tess/bulk\_downloads/bulk\_downloads\_ffi-tp-lc-dv.html}). Part of the data were derived from sources in public domain as given in the respective footnotes. The derived data generated in this research and the code used for the photodynamical analysis will be shared on reasonable request to the corresponding author.



\bibliographystyle{mnras}
\bibliography{example} 




\appendix

\section{Mid-eclipse times of the four eclipsing binaries}

\begin{table*}
\caption{Times of minima of TIC~014839347.}
 \label{Tab:TIC_014839347_ToM}
\begin{tabular}{@{}lrllrllrllrl}
\hline
BJD & Cycle  & std. dev. & BJD & Cycle  & std. dev. & BJD & Cycle  & std. dev. & BJD & Cycle  & std. dev. \\ 
$-2\,400\,000$ & no. &   \multicolumn{1}{c}{$(d)$} & $-2\,400\,000$ & no. &   \multicolumn{1}{c}{$(d)$} & $-2\,400\,000$ & no. &   \multicolumn{1}{c}{$(d)$} & $-2\,400\,000$ & no. &   \multicolumn{1}{c}{$(d)$} \\ 
\hline
58683.46208 &    -0.5 & 0.00036 & 58712.27472 &    24.5 & 0.00042 & 59422.20711 &   640.5 & 0.00022 & 59799.07319 &   967.5 & 0.00018  \\ 
58684.04020 &     0.0 & 0.00009 & 58712.85171 &    25.0 & 0.00013 & 59422.78730 &   641.0 & 0.00008 & 59799.64946 &   968.0 & 0.00011  \\ 
58684.61498 &     0.5 & 0.00123 & 58713.42656 &    25.5 & 0.00023 & 59423.36554 &   641.5 & 0.00029 & 59800.22534 &   968.5 & 0.00025  \\ 
58685.19280 &     1.0 & 0.00020 & 58714.00445 &    26.0 & 0.00015 & 59423.94014 &   642.0 & 0.00010 & 59800.80247 &   969.0 & 0.00010  \\ 
58685.76391 &     1.5 & 0.00036 & 58714.58008 &    26.5 & 0.00055 & 59424.51522 &   642.5 & 0.00027 & 59801.37868 &   969.5 & 0.00025  \\ 
58686.34518 &     2.0 & 0.00019 & 58715.15689 &    27.0 & 0.00015 & 59425.09262 &   643.0 & 0.00013 & 59801.95434 &   970.0 & 0.00008  \\ 
58686.92020 &     2.5 & 0.00063 & 58715.73324 &    27.5 & 0.00083 & 59425.66792 &   643.5 & 0.00026 & 59802.52871 &   970.5 & 0.00025  \\ 
58687.49786 &     3.0 & 0.00028 & 58716.30855 &    28.0 & 0.00022 & 59426.24497 &   644.0 & 0.00015 & 59803.10674 &   971.0 & 0.00012  \\ 
58688.07271 &     3.5 & 0.00060 & 58716.88273 &    28.5 & 0.00017 & 59426.82292 &   644.5 & 0.00034 & 59803.68408 &   971.5 & 0.00028  \\ 
58688.64952 &     4.0 & 0.00012 & 58717.46216 &    29.0 & 0.00017 & 59427.39750 &   645.0 & 0.00013 & 59804.25859 &   972.0 & 0.00010  \\ 
58689.22368 &     4.5 & 0.00056 & 58718.03791 &    29.5 & 0.00072 & 59427.97102 &   645.5 & 0.00047 & 59804.83306 &   972.5 & 0.00026  \\ 
58689.80177 &     5.0 & 0.00023 & 58718.61384 &    30.0 & 0.00021 & 59428.55059 &   646.0 & 0.00011 & 59805.41148 &   973.0 & 0.00008  \\ 
58690.37606 &     5.5 & 0.00056 & 58719.19201 &    30.5 & 0.00026 & 59429.12418 &   646.5 & 0.00041 & 59805.98763 &   973.5 & 0.00032  \\ 
58690.95388 &     6.0 & 0.00014 & 58719.76702 &    31.0 & 0.00034 & 59431.43166 &   648.5 & 0.00018 & 59806.56397 &   974.0 & 0.00008  \\ 
58691.52642 &     6.5 & 0.00044 & 58720.34276 &    31.5 & 0.00021 & 59432.00803 &   649.0 & 0.00008 & 59807.13972 &   974.5 & 0.00034  \\ 
58692.10763 &     7.0 & 0.00021 & 58721.49532 &    32.5 & 0.00057 & 59432.58124 &   649.5 & 0.00161 & 59807.71629 &   975.0 & 0.00013  \\ 
58692.67565 &     7.5 & 0.00170 & 58722.07231 &    33.0 & 0.00011 & 59433.73796 &   650.5 & 0.00120 & 59808.29119 &   975.5 & 0.00023  \\ 
58693.83076 &     8.5 & 0.00055 & 58722.64775 &    33.5 & 0.00078 & 59434.31238 &   651.0 & 0.00008 & 59808.86828 &   976.0 & 0.00010  \\ 
58694.41220 &     9.0 & 0.00016 & 58723.22456 &    34.0 & 0.00026 & 59434.88859 &   651.5 & 0.00034 & 59809.44182 &   976.5 & 0.00039  \\ 
58694.98602 &     9.5 & 0.00039 & 58723.80602 &    34.5 & 0.00164 & 59435.46581 &   652.0 & 0.00009 & 59811.17408 &   978.0 & 0.00009  \\ 
58695.56427 &    10.0 & 0.00012 & 58725.52948 &    36.0 & 0.00020 & 59436.04049 &   652.5 & 0.00027 & 59811.74866 &   978.5 & 0.00028  \\ 
58696.13822 &    10.5 & 0.00063 & 58726.10307 &    36.5 & 0.00038 & 59436.61758 &   653.0 & 0.00009 & 59812.32618 &   979.0 & 0.00009  \\ 
58697.86901 &    12.0 & 0.00018 & 58726.68231 &    37.0 & 0.00010 & 59437.19277 &   653.5 & 0.00041 & 59812.90057 &   979.5 & 0.00029  \\ 
58698.44024 &    12.5 & 0.00036 & 58727.25884 &    37.5 & 0.00021 & 59437.76986 &   654.0 & 0.00010 & 59813.47942 &   980.0 & 0.00010  \\ 
58699.02193 &    13.0 & 0.00017 & 58727.83522 &    38.0 & 0.00020 & 59438.34494 &   654.5 & 0.00028 & 59814.05354 &   980.5 & 0.00015  \\ 
58699.59538 &    13.5 & 0.00079 & 58728.41084 &    38.5 & 0.00022 & 59438.92271 &   655.0 & 0.00013 & 59814.63170 &   981.0 & 0.00012  \\ 
58700.17426 &    14.0 & 0.00015 & 58728.98760 &    39.0 & 0.00012 & 59439.49674 &   655.5 & 0.00031 & 59815.20384 &   981.5 & 0.00035  \\ 
58700.75026 &    14.5 & 0.00052 & 58729.56338 &    39.5 & 0.00037 & 59440.07554 &   656.0 & 0.00021 & 59815.78414 &   982.0 & 0.00010  \\ 
58701.32582 &    15.0 & 0.00015 & 58730.14071 &    40.0 & 0.00022 & 59440.65165 &   656.5 & 0.00023 & 59816.35846 &   982.5 & 0.00018  \\ 
58701.90006 &    15.5 & 0.00019 & 58730.71633 &    40.5 & 0.00038 & 59441.22686 &   657.0 & 0.00011 & 59816.93536 &   983.0 & 0.00011  \\ 
58702.47884 &    16.0 & 0.00018 & 58731.29283 &    41.0 & 0.00014 & 59441.80305 &   657.5 & 0.00041 & 59817.51193 &   983.5 & 0.00026  \\ 
58703.05672 &    16.5 & 0.00099 & 58731.86939 &    41.5 & 0.00035 & 59442.37988 &   658.0 & 0.00011 & 59818.08789 &   984.0 & 0.00010  \\ 
58704.20398 &    17.5 & 0.00257 & 58732.44572 &    42.0 & 0.00014 & 59442.95537 &   658.5 & 0.00028 & 59818.66672 &   984.5 & 0.00022  \\ 
58704.78336 &    18.0 & 0.00015 & 58733.02129 &    42.5 & 0.00023 & 59443.53204 &   659.0 & 0.00013 & 59819.24122 &   985.0 & 0.00011  \\ 
58705.35753 &    18.5 & 0.00045 & 58733.59824 &    43.0 & 0.00026 & 59444.10607 &   659.5 & 0.00036 & 59819.81796 &   985.5 & 0.00026  \\ 
58705.93572 &    19.0 & 0.00020 & 58734.17401 &    43.5 & 0.00028 & 59444.68495 &   660.0 & 0.00009 & 59820.39417 &   986.0 & 0.00011  \\ 
58706.50396 &    19.5 & 0.00117 & 58735.32918 &    44.5 & 0.00115 & 59445.25979 &   660.5 & 0.00027 & 59820.96853 &   986.5 & 0.00023  \\ 
58707.67101 &    20.5 & 0.00074 & 58735.90361 &    45.0 & 0.00020 & 59445.83675 &   661.0 & 0.00011 & 59821.54589 &   987.0 & 0.00009  \\ 
58708.24049 &    21.0 & 0.00018 & 58736.47388 &    45.5 & 0.00151 & 59446.41231 &   661.5 & 0.00025 & 59822.12237 &   987.5 & 0.00036  \\ 
58708.81818 &    21.5 & 0.00024 & 59420.48162 &   639.0 & 0.00012 & 59797.34480 &   966.0 & 0.00009 & 59822.69894 &   988.0 & 0.00010  \\ 
58709.39326 &    22.0 & 0.00023 & 59421.05827 &   639.5 & 0.00022 & 59797.91970 &   966.5 & 0.00024 & 59823.27240 &   988.5 & 0.00021  \\ 
58709.96928 &    22.5 & 0.00051 & 59421.63472 &   640.0 & 0.00013 & 59798.49733 &   967.0 & 0.00012 & 59823.85117 &   989.0 & 0.00010  \\ 
58711.69848 &    24.0 & 0.00020  \\ 
\hline
\end{tabular}
\end{table*}

\begin{table*}
\caption{Times of minima of TIC~066893949.}
 \label{Tab:TIC_066893949_ToM}
\begin{tabular}{@{}lrllrllrllrl}
\hline
BJD & Cycle  & std. dev. & BJD & Cycle  & std. dev. & BJD & Cycle  & std. dev. & BJD & Cycle  & std. dev. \\ 
$-2\,400\,000$ & no. &   \multicolumn{1}{c}{$(d)$} & $-2\,400\,000$ & no. &   \multicolumn{1}{c}{$(d)$} & $-2\,400\,000$ & no. &   \multicolumn{1}{c}{$(d)$} & $-2\,400\,000$ & no. &   \multicolumn{1}{c}{$(d)$} \\ 
\hline
58711.83201 &    -0.5 & 0.00072 & 58733.46997 &     4.0 & 0.00018 & 59435.09934 &   150.0 & 0.00007 & 59802.71738 &   226.5 & 0.00018  \\ 
58714.24737 &     0.0 & 0.00019 & 58735.85916 &     4.5 & 0.00032 & 59437.48807 &   150.5 & 0.00018 & 59805.13517 &   227.0 & 0.00009  \\ 
58716.63577 &     0.5 & 0.00183 & 59420.68260 &   147.0 & 0.00009 & 59439.90506 &   151.0 & 0.00008 & 59812.33065 &   228.5 & 0.00020  \\ 
58719.05333 &     1.0 & 0.00014 & 59423.06986 &   147.5 & 0.00019 & 59442.29382 &   151.5 & 0.00021 & 59814.74714 &   229.0 & 0.00007  \\ 
58721.44275 &     1.5 & 0.00038 & 59425.48859 &   148.0 & 0.00008 & 59444.71088 &   152.0 & 0.00009 & 59817.13492 &   229.5 & 0.00022  \\ 
58726.24731 &     2.5 & 0.00061 & 59427.87673 &   148.5 & 0.00015 & 59797.91148 &   225.5 & 0.00025 & 59819.55200 &   230.0 & 0.00010  \\ 
58728.66411 &     3.0 & 0.00016 & 59430.29490 &   149.0 & 0.00009 & 59800.32911 &   226.0 & 0.00007 & 59821.93961 &   230.5 & 0.00027  \\ 
58731.05216 &     3.5 & 0.00087  \\ 
\hline
\end{tabular}
\end{table*}

\begin{table*}
\caption{Times of minima of TIC~088206187.}
 \label{Tab:TIC_088206187_ToM}
\begin{tabular}{@{}lrllrllrllrl}
\hline
BJD & Cycle  & std. dev. & BJD & Cycle  & std. dev. & BJD & Cycle  & std. dev. & BJD & Cycle  & std. dev. \\ 
$-2\,400\,000$ & no. &   \multicolumn{1}{c}{$(d)$} & $-2\,400\,000$ & no. &   \multicolumn{1}{c}{$(d)$} & $-2\,400\,000$ & no. &   \multicolumn{1}{c}{$(d)$} & $-2\,400\,000$ & no. &   \multicolumn{1}{c}{$(d)$} \\ 
\hline
58816.34459 &    -0.5 & 0.00033 & 58830.55929 &    11.5 & 0.00067 & 59914.20263 &   926.5 & 0.00017 & 59925.45387 &   936.0 & 0.00013  \\ 
58816.93757 &     0.0 & 0.00079 & 58831.15198 &    12.0 & 0.00024 & 59914.79520 &   927.0 & 0.00011 & 59926.04698 &   936.5 & 0.00017  \\ 
58817.53133 &     0.5 & 0.00045 & 58831.74569 &    12.5 & 0.00050 & 59915.38830 &   927.5 & 0.00015 & 59926.63933 &   937.0 & 0.00013  \\ 
58818.12388 &     1.0 & 0.00049 & 58832.33502 &    13.0 & 0.00017 & 59915.97845 &   928.0 & 0.00012 & 59927.23320 &   937.5 & 0.00019  \\ 
58818.71385 &     1.5 & 0.00093 & 58832.92766 &    13.5 & 0.00169 & 59916.57213 &   928.5 & 0.00020 & 59927.82292 &   938.0 & 0.00015  \\ 
58819.30876 &     2.0 & 0.00018 & 58833.51975 &    14.0 & 0.00029 & 59917.16356 &   929.0 & 0.00012 & 59928.41569 &   938.5 & 0.00020  \\ 
58819.90018 &     2.5 & 0.00084 & 58834.11457 &    14.5 & 0.00110 & 59917.75583 &   929.5 & 0.00020 & 59929.00834 &   939.0 & 0.00010  \\ 
58820.49284 &     3.0 & 0.00025 & 58834.70346 &    15.0 & 0.00013 & 59918.34719 &   930.0 & 0.00010 & 59929.60009 &   939.5 & 0.00027  \\ 
58821.08425 &     3.5 & 0.00207 & 58835.29747 &    15.5 & 0.00130 & 59918.93905 &   930.5 & 0.00018 & 59930.19175 &   940.0 & 0.00013  \\ 
58821.67783 &     4.0 & 0.00035 & 58835.88897 &    16.0 & 0.00017 & 59919.53184 &   931.0 & 0.00010 & 59930.78609 &   940.5 & 0.00013  \\ 
58822.26863 &     4.5 & 0.00045 & 58836.48139 &    16.5 & 0.00042 & 59920.12605 &   931.5 & 0.00022 & 59931.37704 &   941.0 & 0.00012  \\ 
58822.86037 &     5.0 & 0.00021 & 58837.07319 &    17.0 & 0.00038 & 59920.71698 &   932.0 & 0.00012 & 59931.97030 &   941.5 & 0.00014  \\ 
58823.45424 &     5.5 & 0.00038 & 58837.66618 &    17.5 & 0.00029 & 59921.30883 &   932.5 & 0.00022 & 59932.56131 &   942.0 & 0.00008  \\ 
58824.04631 &     6.0 & 0.00062 & 58838.25782 &    18.0 & 0.00086 & 59921.90097 &   933.0 & 0.00009 & 59933.15368 &   942.5 & 0.00017  \\ 
58824.63694 &     6.5 & 0.00056 & 58838.84976 &    18.5 & 0.00033 & 59922.49323 &   933.5 & 0.00016 & 59933.74530 &   943.0 & 0.00012  \\ 
58825.22950 &     7.0 & 0.00024 & 58839.44051 &    19.0 & 0.00072 & 59923.08616 &   934.0 & 0.00015 & 59934.33831 &   943.5 & 0.00043  \\ 
58825.82227 &     7.5 & 0.00021 & 58840.03120 &    19.5 & 0.00044 & 59924.27054 &   935.0 & 0.00015 & 59934.92946 &   944.0 & 0.00014  \\ 
58826.41376 &     8.0 & 0.00018 & 58840.62481 &    20.0 & 0.00047 & 59924.86028 &   935.5 & 0.00025 & 59935.52280 &   944.5 & 0.00025  \\ 
58827.00838 &     8.5 & 0.00043 & 59913.61059 &   926.0 & 0.00013  \\ 
\hline
\end{tabular}
\end{table*}

\begin{table*}
\caption{Times of minima of TIC~298714297.}
 \label{Tab:TIC_298714297_ToM}
\begin{tabular}{@{}lrllrllrllrl}
\hline
BJD & Cycle  & std. dev. & BJD & Cycle  & std. dev. & BJD & Cycle  & std. dev. & BJD & Cycle  & std. dev. \\ 
$-2\,400\,000$ & no. &   \multicolumn{1}{c}{$(d)$} & $-2\,400\,000$ & no. &   \multicolumn{1}{c}{$(d)$} & $-2\,400\,000$ & no. &   \multicolumn{1}{c}{$(d)$} & $-2\,400\,000$ & no. &   \multicolumn{1}{c}{$(d)$} \\ 
\hline
58711.49994 &     0.0 & 0.00037 & 58737.24706 &    24.0 & 0.00024 & 59816.56348 &  1030.0 & 0.00007 & 59835.33955 &  1047.5 & 0.00009  \\ 
58712.03597 &     0.5 & 0.00063 & 59797.25118 &  1012.0 & 0.00005 & 59817.09956 &  1030.5 & 0.00028 & 59835.87540 &  1048.0 & 0.00005  \\ 
58712.57261 &     1.0 & 0.00040 & 59797.78710 &  1012.5 & 0.00022 & 59817.63617 &  1031.0 & 0.00018 & 59836.41296 &  1048.5 & 0.00011  \\ 
58713.64553 &     2.0 & 0.00032 & 59798.32436 &  1013.0 & 0.00010 & 59818.17406 &  1031.5 & 0.00031 & 59836.94845 &  1049.0 & 0.00005  \\ 
58714.18127 &     2.5 & 0.00043 & 59798.86084 &  1013.5 & 0.00022 & 59818.70930 &  1032.0 & 0.00007 & 59837.48540 &  1049.5 & 0.00014  \\ 
58714.71826 &     3.0 & 0.00054 & 59799.93378 &  1014.5 & 0.00038 & 59819.24606 &  1032.5 & 0.00021 & 59838.02128 &  1050.0 & 0.00005  \\ 
58715.79108 &     4.0 & 0.00017 & 59800.46966 &  1015.0 & 0.00015 & 59819.78208 &  1033.0 & 0.00019 & 59839.09430 &  1051.0 & 0.00004  \\ 
58716.32573 &     4.5 & 0.00092 & 59801.00665 &  1015.5 & 0.00028 & 59820.31865 &  1033.5 & 0.00033 & 59839.63180 &  1051.5 & 0.00010  \\ 
58716.86396 &     5.0 & 0.00038 & 59801.54269 &  1016.0 & 0.00007 & 59820.85502 &  1034.0 & 0.00008 & 59840.16712 &  1052.0 & 0.00004  \\ 
58717.93698 &     6.0 & 0.00020 & 59802.07904 &  1016.5 & 0.00047 & 59821.39111 &  1034.5 & 0.00019 & 59840.70440 &  1052.5 & 0.00011  \\ 
58718.47299 &     6.5 & 0.00059 & 59802.61562 &  1017.0 & 0.00015 & 59821.92778 &  1035.0 & 0.00019 & 59841.24016 &  1053.0 & 0.00005  \\ 
58719.00964 &     7.0 & 0.00038 & 59803.15228 &  1017.5 & 0.00020 & 59822.46474 &  1035.5 & 0.00025 & 59841.77669 &  1053.5 & 0.00010  \\ 
58720.08276 &     8.0 & 0.00044 & 59803.68825 &  1018.0 & 0.00009 & 59823.00079 &  1036.0 & 0.00008 & 59842.31323 &  1054.0 & 0.00004  \\ 
58720.61764 &     8.5 & 0.00054 & 59804.76143 &  1019.0 & 0.00010 & 59823.53781 &  1036.5 & 0.00035 & 59842.84990 &  1054.5 & 0.00009  \\ 
58721.15530 &     9.0 & 0.00040 & 59805.29855 &  1019.5 & 0.00022 & 59824.07355 &  1037.0 & 0.00036 & 59843.38607 &  1055.0 & 0.00005  \\ 
58722.22824 &    10.0 & 0.00009 & 59805.83441 &  1020.0 & 0.00013 & 59825.68383 &  1038.5 & 0.00014 & 59843.92310 &  1055.5 & 0.00009  \\ 
58722.76369 &    10.5 & 0.00080 & 59806.37126 &  1020.5 & 0.00031 & 59826.21946 &  1039.0 & 0.00005 & 59844.45912 &  1056.0 & 0.00006  \\ 
58723.30097 &    11.0 & 0.00042 & 59806.90729 &  1021.0 & 0.00014 & 59826.75659 &  1039.5 & 0.00011 & 59844.99635 &  1056.5 & 0.00011  \\ 
58725.44656 &    13.0 & 0.00025 & 59807.44376 &  1021.5 & 0.00027 & 59827.29228 &  1040.0 & 0.00004 & 59845.53190 &  1057.0 & 0.00004  \\ 
58726.51958 &    14.0 & 0.00009 & 59807.98022 &  1022.0 & 0.00005 & 59827.82956 &  1040.5 & 0.00014 & 59846.60491 &  1058.0 & 0.00005  \\ 
58727.05501 &    14.5 & 0.00107 & 59808.51656 &  1022.5 & 0.00020 & 59828.36514 &  1041.0 & 0.00005 & 59847.14169 &  1058.5 & 0.00016  \\ 
58727.59211 &    15.0 & 0.00019 & 59809.05307 &  1023.0 & 0.00018 & 59828.90253 &  1041.5 & 0.00012 & 59847.67775 &  1059.0 & 0.00005  \\ 
58728.66521 &    16.0 & 0.00011 & 59809.58973 &  1023.5 & 0.00030 & 59829.43803 &  1042.0 & 0.00005 & 59848.21495 &  1059.5 & 0.00010  \\ 
58729.20005 &    16.5 & 0.00225 & 59811.19906 &  1025.0 & 0.00014 & 59829.97538 &  1042.5 & 0.00012 & 59848.75079 &  1060.0 & 0.00004  \\ 
58729.73778 &    17.0 & 0.00018 & 59811.73565 &  1025.5 & 0.00027 & 59830.51100 &  1043.0 & 0.00005 & 59849.28744 &  1060.5 & 0.00014  \\ 
58730.81024 &    18.0 & 0.00046 & 59812.27176 &  1026.0 & 0.00008 & 59831.04834 &  1043.5 & 0.00010 & 59849.82361 &  1061.0 & 0.00005  \\ 
58731.34656 &    18.5 & 0.00033 & 59812.80619 &  1026.5 & 0.00019 & 59831.58384 &  1044.0 & 0.00006 & 59850.36021 &  1061.5 & 0.00009  \\ 
58731.88324 &    19.0 & 0.00004 & 59813.34481 &  1027.0 & 0.00016 & 59832.12125 &  1044.5 & 0.00013 & 59850.89676 &  1062.0 & 0.00005  \\ 
58732.95598 &    20.0 & 0.00039 & 59813.88129 &  1027.5 & 0.00019 & 59832.65671 &  1045.0 & 0.00005 & 59851.43322 &  1062.5 & 0.00016  \\ 
58734.02900 &    21.0 & 0.00009 & 59814.41770 &  1028.0 & 0.00007 & 59833.19406 &  1045.5 & 0.00010 & 59851.96958 &  1063.0 & 0.00004  \\ 
58735.10149 &    22.0 & 0.00046 & 59814.95407 &  1028.5 & 0.00022 & 59833.72953 &  1046.0 & 0.00004 & 59852.50629 &  1063.5 & 0.00010  \\ 
58735.63702 &    22.5 & 0.00033 & 59815.49060 &  1029.0 & 0.00020 & 59834.26648 &  1046.5 & 0.00013 & 59853.04255 &  1064.0 & 0.00007  \\ 
58736.17434 &    23.0 & 0.00002 & 59816.02711 &  1029.5 & 0.00026 & 59834.80254 &  1047.0 & 0.00005  \\ 
\hline
\end{tabular}
\end{table*}


\bsp	
\label{lastpage}
\end{document}